\documentstyle[pra,aps,epsf,eqsecnum]{revtex}
\begin{document}

\newcommand\be{\begin{equation}}
\newcommand\ee{\end{equation}}
\newcommand\bea{\begin{eqnarray}}
\newcommand\eea{\end{eqnarray}}
\newcommand\oupb{UPB\ }
\newcommand\eps{\epsilon}
\newcommand{\rank}{\rm rank}
\newcommand\oupbs{UPBs\ }

\newcommand{\ket}[1]{| #1 \rangle}
\newcommand{\bra}[1]{\langle #1 |}
\newcommand{\braket}[2]{\langle #1 | #2 \rangle}
\newcommand{\proj}[1]{| #1\rangle\!\langle #1 |}
\newcommand{\ba}{\begin{array}}
\newcommand{\ea}{\end{array}}

\newtheorem{theo}{Theorem}
\newtheorem{defi}{Definition}
\newtheorem{lem}{Lemma}
\newtheorem{exam}{Example}
\newtheorem{prop}{Property}
\newtheorem{conj}{Conjecture}
\newtheorem{cor}{Corollary}
\newtheorem{propo}{Proposition}
\author{David P. DiVincenzo$^*$, Tal Mor$^\dag$,
Peter W. Shor$^\ddag$, John A. Smolin$^*$ and Barbara M. Terhal$^\S$}
 
\title{Unextendible Product Bases, Uncompletable Product Bases
and Bound Entanglement}
 
\address{\vspace*{1.2ex} \hspace*{0.5ex}{$^*$ IBM T.J. Watson Research
Center, Yorktown Heights, NY 10598, USA, $^\dag$ Dept. of Electrical Engineering, UCLA, Los
Angeles, CA 90095-1594, USA, $^\ddag$AT\&T Research, Florham Park, NJ 07932, USA, $^\S$ ITF, UvA, Valckenierstraat 65, 1018 XE Amsterdam, and CWI,
Kruislaan 413, 1098 SJ Amsterdam, The Netherlands.\\ }}
 
\date{\today}
 
\maketitle

\begin{abstract}
We report new results and generalizations of our work on
unextendible product bases (UPB), uncompletable product bases and bound entanglement.
We present a new construction for bound entangled states based
on product bases which are only completable in a locally
extended Hilbert space. We introduce a very useful representation
of a product basis, an orthogonality graph. Using this representation we give a complete characterization of unextendible
product bases for two qutrits. We present several generalizations of UPBs to arbitrary high dimensions and
multipartite systems. We present a sufficient condition for
sets of orthogonal product states to be distinguishable by
separable superoperators. We prove that bound entangled states
cannot help increase the distillable entanglement
of a state beyond its regularized entanglement of formation assisted by 
bound entanglement.
%
\end{abstract}

\section{Introduction}
\label{intro}
 
One of the essential features of quantum information is its capacity for
entanglement. When pure state entanglement is shared by two or more
parties, it permits them to send quantum data with classical
communication via teleportation \cite{tele}. In a more general
situation two parties may not start with a set of pure entangled
states, but with a noisy quantum channel.  To achieve their goal of
transmitting quantum data over this channel, they could use an error
correcting code \cite{gotthesis}, or alternatively they can attempt to share
entanglement through the channel and later use teleportation.  In the
latter case, the protocol starts with the preparation of entangled
states by, say, Alice, who sends half of each entangled state through
the noisy channel to her partner Bob.  Since the channel is noisy these
states will not directly be useful for teleportation. As a next step
Alice and Bob go through a protocol of purification \cite{distprl};
they try to distill as many as possible pure entangled states out
of the set of noisy ones using only local operations and classical
communication. We will abbreviate such local quantum operations
supplemented by classical communication hereafter as ``LQ+CC'' operations.
Finally, they can use these distilled states to teleport the quantum
data.  The amount of quantum data that can be sent via the protocol of
distillation and teleportation can be {\em higher} than by `direct'
quantum data transmission using error correcting codes
\cite{bdsw}.  This has been one of the motivations for studying
bipartite mixed state entanglement.
 
Let us review the definition of entanglement and introduce some notation.
A density matrix $\rho$ on a multipartite Hilbert space ${\cal H}$ is
separable if we can find a decomposition of $\rho$ into an
ensemble of pure product states in ${\cal H}$. Thus, for a bipartite
Hilbert space a separable density matrix $\rho$ can always be written
as
\be
\rho=\sum_i p_i \ket{\alpha_i}\bra{\alpha_i} \otimes \ket{\beta_i} \bra{\beta_i},
\ee
where $p_i \geq 0$. When a density matrix is not separable, the density matrix is called entangled.
In the following we use the notation $n \otimes m$ or ${\cal H}_n \otimes {\cal H}_m$ to denote the
tensor product between a $n$-dimensional Hilbert space and a
$m$-dimensional Hilbert space. A Trace-preserving Completely Positive
linear map ${\cal S}$ is abbreviated as a TCP map ${\cal S}$.
When a Hermitian matrix $\sigma$ has eigenvalues greater than or equal to zero,
we denote this as $\sigma \geq 0$, i.e. $\sigma$ is a positive semidefinite
matrix. We denote the set of linear operators on a Hilbert space ${\cal H}$ as
$B({\cal H})$.
 
The theory of positive linear maps has turned out to be an important
tool in characterizing bipartite mixed state entanglement \cite{nec_horo}.
It has been shown \cite{pptnodist} that all density matrices $\rho$ on
${\cal H}_n \otimes {\cal H}_m$ which remain positive semidefinite under the
partial transposition (PT) map, i.e. $({\bf 1} \otimes T)(\rho) \geq 0$, where
$T$ is matrix transposition \cite{fnt1}, are not distillable.  We will say that
such density matrices have the PPT property or ``are PPT''.
Here a density matrix $\rho$ is called distillable when for all $\eps > 0$, 
there exists an integer $n$ and a LQ+CC procedure ${\cal S}:B({\cal H}^{\otimes n} \rightarrow {\cal H}_2$ with 
$\bra{\Psi^-}|{\cal S}(\rho^n)\ket{\Psi^-} \geq 1-\eps$, where $\ket{\Psi^-}$ 
is a singlet state. 

A state which has entanglement but which is not distillable is called
a {\em bound entangled} state.  All entangled states which are PPT
are thus bound entangled.  But do such states exist?  It was shown in
Ref. \cite{nec_horo} that entangled states with the PPT property do not
exist in Hilbert spaces $2 \otimes 2$ and $2 \otimes 3$. The first
examples of entangled density matrices with the PPT property in
higher dimensional Hilbert spaces were found by P. Horodecki \cite{bepawel}.  In Ref. \cite{upb1} we presented the
first {\em method} for constructing bound entangled PPT states. This method
relies on the notion of an {\em unextendible product basis} or UPB.
This construction has also led to a method for constructing indecomposable
positive linear maps \cite{terhalposmap}. In Ref. \cite{upb1} we have given
several examples of unextendible product bases, and therefore of bound
entangled states. We showed that the notion of an unextendible product
basis has another interesting feature, namely the states in the
unextendible product basis are not exactly distinguishable by local
quantum operations and classical communication. They form a demonstration of
the phenomenon of ``nonlocality without entanglement'' \cite{qne}.
 
In this paper we continue the work that was started in Ref. \cite{upb1}.
We will review many of the results that were presented in Ref. \cite{upb1}. The paper is organized in the following way.
 
In section \ref{properties} we review some of the definitions and results that were presented in Ref. \cite{upb1}.
In section \ref{ucpb} we present a first example and indicate a method
to make bound entangled states which are based on uncompletable but not
strongly uncompletable product bases.
In section \ref{usesepsup} we present a sufficient condition for
members of an orthogonal product basis to be distinguishable by separable
superoperators.
In section \ref{orthograph} we introduce the notion of an orthogonality
graph associated with a product basis; this notion helps us in
establishing a complete characterization of all unextendible product
bases in $3 \otimes 3$. In section \ref{manyexam} we
present unextendible product bases for multipartite
and bipartite high dimensional Hilbert spaces. Again we will make fruitful use
of the notion of an orthogonality graph.
In section \ref{usebe} we report several results that are obtained
in considering the use of bound entangled states. We will prove a 
restriction on the use of bound entangled states in the distillation of
entangled states. In section \ref{binding} we relate the sharing of
bound entanglement to the possession of a quantum channel, namely
a binding entanglement channel.

\section{Properties of Uncompletable and Unextendible Product Bases}
\label{properties}
 
In this section we exhibit various properties of uncompletable
and unextendible product bases, and explore
their relation to local distinguishability of sets of product states
and bound entanglement.

\subsection{Definitions and Counting Lemma}
\label{defin}
 
We give the definitions of three kinds of sets of
orthogonal product states. First we define an unextendible and
an uncompletable product basis:
 
\begin{defi}
Consider a multipartite quantum system ${\cal H}=\bigotimes_{i=1}^m
{\cal H}_i$ with $m$ parties. An orthogonal product basis (PB) is a set {\rm S} of pure
orthogonal product states spanning a subspace ${\cal H}_{\rm S}$ of
${\cal H}$. An uncompletable product basis (UCPB) is a PB whose
complementary subspace ${\cal H}_{\rm S}^{\perp}$, i.e. the subspace
 in ${\cal H}$ spanned by vectors that are orthogonal to all the vectors
in ${\cal H}_{\rm S}$, contains fewer mutually orthogonal product
states than its dimension. An unextendible product basis (UPB) is an
uncompletable product basis for which ${\cal H}_{\rm S}^{\perp}$ contains no product state.
\end{defi}
 
Thus, for an unextendible product basis
S, it is not possible to find a product vector in ${\cal H}$ that is
orthogonal to all the members in S. For an uncompletable product
basis S, it may be possible to find product vectors that are orthogonal to
all the states in S, however, we will never be able to find enough states
so as to complete the set S to a full basis for ${\cal H}$.
 
Now we give the next definition, that of a strongly uncompletable
product basis, for which we will use the notion of a locally extended
Hilbert space. Let ${\cal H}=\bigotimes_{i=1}^m {\cal H}_i$, a Hilbert space of an
$m$-partite system. A locally extended Hilbert space is defined
as ${\cal H}_{ext}=\bigotimes_{i=1}^m ({\cal H}_i \oplus
{\cal H}_i')$, where ${\cal H}_i'$ is a local extension.
When we are given a set of states in ${\cal H}$ we can consider
properties of this set embedded in a locally extended Hilbert space
${\cal H}_{ext}$.
 
\begin{defi}
Consider a multipartite quantum system ${\cal H}=\bigotimes_{i=1}^m
{\cal H}_i$ with $m$ parties. A strongly uncompletable product
basis (SUCPB) is a PB spanning a subspace ${\cal H}_{{\rm S}}$ in
a locally extended Hilbert space ${\cal H}_{ext}$ such that for all
${\cal H}_{ext}$ the subspace ${\cal H}_{{\rm S}}^{\perp}$
(${\cal H}_{ext}={\cal H}_{\rm S} \oplus {\cal H}_{{\rm S}}^{\perp}$)
contains fewer mutually orthogonal product states than its dimension.
\end{defi}
 
Thus a strongly uncompletable product basis cannot be completed to a full
product basis of some extended Hilbert space ${\cal H}_{ext}$. In section
\ref{ucpb} we will give an example of a PB that is uncompletable but not
strongly uncompletable.
 
\begin{figure}
\epsfxsize=3.0in
\centering
\leavevmode\epsfbox{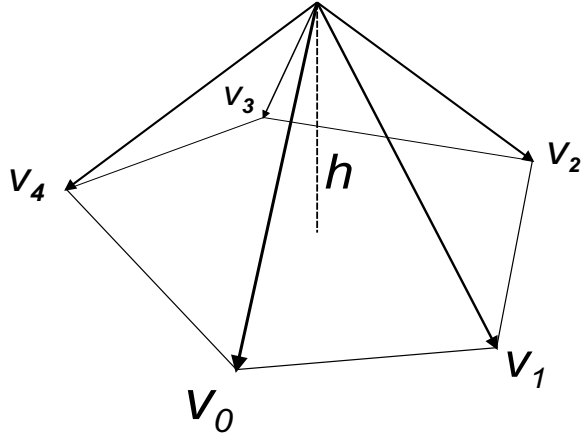}
\caption{${\bf Pyramid}$ vectors in real 3-space.  The height $h$ is chosen
so that
$\vec{v}_0 \perp \vec{v}_{2,3}$ etc.}
\label{Pyramid}
\end{figure}

We will review an example of an unextendible
product basis of five states in $3 \otimes 3$ (two qutrits) given in Ref. \cite{upb1}.
Let $\vec{v}_0,\vec{v}_1,\ldots,\vec{v}_4$ be five vectors in real
three dimensional space forming the apex of
a regular pentagonal pyramid, the height $h$ of the pyramid being chosen
such that nonadjacent vectors are orthogonal (see Fig. \ref{Pyramid}).
The vectors are
\be
\vec{v}_i=N(\cos{{ 2 \pi i \over 5}},\sin{{2 \pi i}\over 5},h),\;\;
i=0,\ldots,4,
\label{defP}
\ee
with $h={1 \over 2} \sqrt{1+\sqrt{5}}$ and $N=2/\sqrt{5+\sqrt{5}}$.
Then the following five states in a $3 \otimes 3$ Hilbert space
form a UPB, henceforth denoted {\bf Pyramid}
\be
\vec{\psi_i}= \vec{v}_i \otimes \vec{v}_{2i \bmod 5}, \;\; i=0,\ldots,4.
\ee

 
To see that these five states form a UPB, note first that they are
mutually orthogonal: states whose indices differ by 2 mod 5 are
orthogonal for the first party (``Alice''); those whose indices differ by
1 mod 5 are orthogonal for the second party (``Bob''). For a new state
to be orthogonal to all the existing ones, it would have to be
orthogonal to at least three of Alice's states or at least three of
Bob's states. However this is impossible, since any set of three vectors
$\vec{v}_i$ spans the full three dimensional space.  Therefore
the entire four dimensional subspace ${\cal H}_{\bf Pyramid}^{\perp}$ contains
no product state.
 
We formalize this observation by giving the necessary and sufficient
condition for extendibility of a PB (the proof is given in Ref. \cite{upb1}):
 
\begin{lem} \cite{upb1} Let ${\rm S}=\{(\psi_j\!\equiv\!\bigotimes_{i=1}^m\varphi_{i,j}):j=1\ldots n\}$ be an orthogonal
product basis (PB) spanning a subspace of the Hilbert space of an
$m$-partite quantum system ${\cal H}=\bigotimes_{i=1}^m
{\cal H}_i$ with $\dim {\cal H}_i=d_i$.  Let $\rm P$ be a partition of $\rm S$ into a
number $m$ of disjoint subsets equal to the number of parties:
${\rm S}={\rm S}_1\cup {\rm S}_2\cup \ldots {\rm S}_m$.
Let $r_i=\rank\{{\it \varphi_{i,j}:\psi_j\in {\rm S}_i}\}$
be the local rank of subset ${\rm S}_i$ as
seen by the $i$th party.  Then $\rm S$ is extendible if and only if there
exists a partition $\rm P$ such that for all $i=1\ldots m$, the local rank
of the $i$th subset is less than the dimensionality of the $i$th
party's Hilbert space.  That is to say, $\rm S$ is extendible iff
$\exists_{\rm P}\forall_i \; r_i<d_i$.
\label{rule1}
\end{lem}
 
The lemma provides a simple lower bound on the number of states $n$ in a
UPB,
\be
n \geq \sum_i (d_i-1)+1,
\label{min}
\ee
since, for smaller $n$, one can partition S into sets of size $|{\rm S}_i|
\leq d_i-1$ and thus $r_i < d_i$ for all $m$ parties.

\subsection{Unextendible Product Bases and Bound Entanglement}
\label{upbtobe}
 
Every UPB on a bipartite or multipartite Hilbert space gives rise
to a bound entangled state which has the PPT property. The construction
is the following:
 
\begin{theo}\cite{upb1}
Let ${\rm S}$ be a UPB $\{\psi_i: i=1,\ldots,n\}$ in a Hilbert space of total dimension $D$. The density matrix $\bar{\rho}$ that is proportional to the projector onto ${\cal H}_{\rm S}^{\perp}$,
\be
\bar{\rho}= {1 \over D-n}\left({\bf 1} -\sum_{j=1}^n \proj{\psi_j}\right),
\label{compl}
\ee
is a bound entangled density matrix.
\label{rhobar}
\end{theo}
 
\noindent {\it Proof}. By definition, ${\cal H}_{\rm S}^{\perp}$ contains
no product states.  Therefore $\bar\rho$ is entangled.  If the UPB is a bipartite UPB then we can directly apply the PT map to
$\bar{\rho}$ and find that $({\bf 1} \otimes T)(\bar{\rho}) \geq 0$.
Then we use the fact from Ref. \cite{pptnodist} that if a bipartite density
matrix has the PPT property, it is not distillable. To derive the PPT property of $\bar{\rho}$ we recall that the PT map is linear so we may apply it separately to the identity and to the projector onto ${\cal H}_{\rm S}$ in $\bar{\rho}$.
The identity is invariant under the PT map.  Each projector onto a
product state is of the form $\proj{\psi_A} \otimes \proj{\psi_B}$ and as
such will be mapped onto
\bea
({\bf 1} \otimes T)(\proj{\psi_A} \otimes \proj{\psi_B})=\proj{\psi_A} \otimes T(\proj{\psi_B}) \nonumber \\
= \proj{\psi_A} \otimes \proj{\psi_B^*}.
\label{projmap}
\eea
The product states making up the UPB are mapped onto another set of orthogonal product states. Therefore $({\bf 1} \otimes T)(\bar{\rho})\geq 0$.
 
In case of a multipartite UPB the PPT condition cannot be used directly.
However we can use the above argument to show
that under every bipartite partitioning of the parties $\bar{\rho}$ is PPT.  Thus no
entanglement can be distilled across any bipartite cut.  If any
pure `global' entanglement could be distilled it could be used to
create entanglement across a bipartite cut. Therefore no entanglement
can be distilled and thus the density matrix $\bar\rho$ is bound entangled.
$\Box$
 
It was pointed out by C.H. Bennett \cite{charliepriv} that it is a
simple matter to create a set S of nonorthogonal product states in a Hilbert
space ${\cal H}$ such that no other product state can be found in ${\cal H}$
that is orthogonal to all the states in S. In fact, except for a set of measure zero, any set of randomly
chosen product vectors whose number satisfies Eq. (\ref{min}) will be
unextendible in this sense.  For every partitioning the new product vector to be added to
the set will have to be orthogonal to $d_i$ other vectors for at least one
party $i$. However $d_i$ randomly chosen vectors will typically
span a $d_i$-dimensional space, and therefore such a new product vector that
is orthogonal to all the members in the set cannot exist. It is not clear how such a nonorthogonal set of product states
could lead to a bound entangled state. The projector on
${\cal H}_{S}^{\perp}$ where ${\cal H}_{\rm S}$ is now the space
spanned by the nonorthogonal product vectors, is entangled, but
does not necessarily have the PPT property; in the
set of orthogonal vectors obtained by Gram-Schmidt orthogonalization of the
nonorthogonal product vectors we might find entangled vectors and
therefore $\bar{\rho}$ might not have the PPT property.

\subsection{Local Distinguishability of Product Bases}
\label{locdist}
 
When two parties Alice and Bob possess one state out of an
ensemble of orthogonal product states, we may ask whether it is
possible for them to determine exactly which state they have by
performing local quantum operations and classical communication.
As the states are orthogonal, a joint measurement for Alice and Bob
that exactly distinguishes the states is always possible.
 
In Ref. \cite{upb1} we found that when a set of product states in a multipartite Hilbert space is strongly uncompletable, it implies that the members in the set cannot be distinguished by LQ+CC. This result is captured in the following
lemma:
 
\begin{lem}\cite{upb1}
Given a set $\rm S$ of orthogonal product states on ${\cal
H}=\bigotimes_{i=1}^m {\cal H}_i$. If the set $\rm S$ is exactly distinguishable
by local von Neumann measurements and classical communication then it
is completable in $\cal H$.  If $\rm S$ is exactly distinguishable by local
POVMs and classical communication then the set can be completed in
some extended Hilbert space ${\cal H'}=\bigotimes_{i=1}^m ({\cal H}_i \oplus
{\cal H}_i')$.
\label{immeasurability}
\end{lem}
 
The proof is given in Ref. \cite{upb1}. We note that in the lemma we only
allow POVMs with a finite number of
outcomes and we only allow a finite number of rounds of POVM measurements.
This restriction comes about because we use Neumark's theorem \cite{peresbook} to
convert a POVM measurement by a party $i$ into a von Neumann measurement on a
locally extended Hilbert space ${\cal H}_{i,ext}={\cal H}_i \oplus
{\cal H}_i'$. When the number of POVM measurement outcomes is infinite,
then the extended Hilbert space is infinite dimensional. It is not
clear how one can speak of completing a set of states to a full product
basis for an infinite dimensional Hilbert space. We avoid the same problem
by excluding the possibility for an infinite number of rounds of
POVM measurements.
 
It is possible to strengthen the lemma one step further and include
measurements that do not exactly distinguish the set of states, but
make an arbitrary small error $\eps$. By this we mean that we use
a measurement and a decision scheme for which the probability of
correctly deciding what state the parties were given, is greater than or equal to $1-\eps$ for all possible
states that the parties can possess. When we allow only LQ+CC by finite
means, both in space and time, it is possible to prove that when measurement
plus decision schemes exist that make an arbitrary small
error $\eps$ for all $\eps > 0$, then there will also exist a scheme that makes no error. The proof of this result is given in Ref. \cite{thesisbmt} and relies on the fact
that the set of measurement and decision schemes is a finite union of
compact sets.
 
We would like to stress that the converse of Lemma \ref{immeasurability}
does not hold. There do exist sets of orthogonal product states that
are not distinguishable by LQ+CC, but which are completable. A prime
example is the set of states given in Ref. \cite{qne}. For this set it was
proved that even by allowing an infinite number of rounds of local measurements, it was
not possible to distinguish the members with arbitrary small probability
of error.
 
\subsection{Uncompletable Product Bases and Bound Entanglement}
\label{ucpb}
 
Lemma \ref{immeasurability} relates uncompletable product bases
(UCPBs and SUCPBs) to distinguishability. We may also ask how these (S)UCPBs
relate to bound entanglement. First we recall a simple observation
that was presented in Ref. \cite{upb1}:
 
\begin{propo}\cite{upb1}
Given a PB {\rm S} on ${\cal H}=\bigotimes_{i=1}^m {\cal H}_i$. If the set
{\rm S} is completable in ${\cal  H}$ or a locally extended Hilbert space
${\cal H}_{ext}$, then the density matrix $\bar{\rho}_{\rm S}$ is separable.
\label{cosep}
\end{propo}
 
This directly implies that a UPB is strongly uncompletable, since the state
$\bar{\rho}$ corresponding to the UPB is always entangled.
 
We now ask what properties the projector onto ${\cal H}_{\rm S}^{\perp}$ has
when S is a UCPB or a SUCPB. Certainly, this projector has the PPT property; it will
thus either be separable or have bound entanglement. In order to explore this question, we return to an example that was given
in Ref. \cite{upb1}. We consider the PB {\bf Pyr34}, a curious set of states
in $3 \otimes 4$ of which the members are distinguishable by local POVMs
and classical communication, but {\em not} by von Neumann measurements. {\bf Pyr34} consists of the states $\vec{v}_j\otimes \vec{w}_j,\;\; j=0,\ldots,4$ with $\vec{v}_j$ the states of the {\bf Pyramid} \oupb as in Eq. (\ref{defP}) and
$\vec{w}_j$ defined as
\begin{eqnarray}
\nonumber\vec{w}_j&=N(\sqrt{\cos(\pi/5)} \cos(2j\pi/5),\sqrt{\cos(\pi/5)} \sin(2j\pi/5),\\
&\sqrt{\cos(2\pi/5)}\cos(4j\pi/5),\sqrt{\cos(2\pi/5)}\sin(4j\pi/5)),
\end{eqnarray}
with normalization $N=\sqrt{2/\sqrt{5}}$. Note that $\vec{w}_j^T
\vec{w}_{j+1}=0$ (addition mod $5$). One can show that this set,
albeit extendible on $3 \otimes 4$, is not {\it completable}: One
can at most add three vectors like $\vec{v}_0 \otimes
(\vec{w}_0,\vec{w}_1,\vec{w}_4)^{\perp}$, $\vec{v}_3 \otimes
(\vec{w}_2,\vec{w}_3,\vec{w}_4)^{\perp}$ and
$(\vec{v}_0,\vec{v}_3)^{\perp} \otimes
(\vec{w}_1,\vec{w}_2,\vec{w}_4)^{\perp}$. Therefore this set is an example
of a UCPB. However it is possible to distinguish the members of this
set by local POVM measurements and classical communication. With this
property, Lemma \ref{immeasurability} implies that the set is completable
in a locally extended Hilbert space. The set is thus {\em not} strongly
uncompletable. This again implies with Proposition \ref{cosep} that the state
$\bar{\rho}_{{\bf Pyr34}}$ is a separable density matrix.
 
The local POVM that distinguishes the members of {\bf Pyr34} starts with a POVM
performed by Bob on the four-dimensional side. Bob's POVM
has five projector elements, each projecting onto a vector
$\vec{u}_j=N(-\sin(2 j \pi/5),\linebreak \cos(2 j \pi/5),-\sin(4 j \pi/5),\cos(4j
\pi/5))$ with $j=0,\ldots,4$, and normalization $N=1/\sqrt{2}$.  Note
that $\vec{u}_0$ is orthogonal to vectors $\vec{w}_0,\vec{w}_2$ and
$\vec{w}_3$, or, in general, $\vec{u}_i$ is orthogonal to
$\vec{w}_i,\vec{w}_{i+2},\vec{w}_{i+3}$ (addition mod $5$). This means
that when Bob obtains his POVM measurement outcome, three vectors are excluded
from the set; then the remaining two vectors on Alice's side,
$\vec{v}_{i+1}$ and $\vec{v}_{i+4}$, are orthogonal and can thus be
distinguished.
 
The completion of the {\bf Pyr34} set is particularly
simple: Bob's Hilbert space is extended to a five dimensional space.
The POVM measurement can be extended as a projection measurement in
this five-dimensional space with orthogonal projections onto the states
$\vec{x}_i=(\vec{u}_i,0)+\frac{1}{2}(0,0,0,0,1)$.
Then a completion of the set in $3 \otimes 5$ are the following ten states:
\be
\ba{lr} (\vec{v}_1,\vec{v}_4)^{\perp} \otimes \vec{x}_0, & \vec{v}_0
\otimes (\vec{w}_0^{\perp} \in \mbox{span}(\vec{x}_4,\vec{x}_1)), \\
(\vec{v}_0,\vec{v}_2)^{\perp} \otimes \vec{x}_1, & \vec{v}_1 \otimes
(\vec{w}_1^{\perp} \in \mbox{span}(\vec{x}_0,\vec{x}_2)), \\
(\vec{v}_1,\vec{v}_3)^{\perp} \otimes \vec{x}_2, & \vec{v}_2 \otimes
(\vec{w}_2^{\perp} \in \mbox{span}(\vec{x}_1,\vec{x}_3)), \\
(\vec{v}_2,\vec{v}_4)^{\perp} \otimes \vec{x}_3, & \vec{v}_3 \otimes
(\vec{w}_3^{\perp} \in \mbox{span}(\vec{x}_2,\vec{x}_4)), \\
(\vec{v}_0,\vec{v}_3)^{\perp} \otimes \vec{x}_4, & \vec{v}_4 \otimes
(\vec{w}_4^{\perp} \in \mbox{span}(\vec{x}_3,\vec{x}_0)).  \ea \ee
Because the set {\bf Pyr34} is uncompletable in the Hilbert space $3 \otimes 4$,
the state $\bar{\rho}_{{\bf Pyr34}}$ has the notable property that although
it is separable, it is not decomposable using orthogonal product states \cite{barely}. If it were, those states would form a completion of the set {\bf Pyr34}.
 
Let us now take the set {\bf Pyr34} and add one product state, say the vector
\be
\vec{v}_0 \otimes (\vec{w}_0,\vec{w}_1,\vec{w}_4)^{\perp},
\ee
to make it a six-state ensemble {\bf Pyr34}$^+$. The density matrix $\bar{\rho}_{{\bf Pyr34}^+}$ has rank $12-6=6$.
Is $\bar{\rho}_{{\bf Pyr34}^+}$ still a separable
density matrix? We can enumerate the product states that are orthogonal to the
members of {\bf Pyr34}$^+$, which are not all mutually orthogonal:
\be
\ba{r}
\vec{v}_3 \otimes (\vec{w}_2,\vec{w}_3,\vec{w}_4)^{\perp}, \\
\vec{v}_2 \otimes (\vec{w}_1,\vec{w}_2,\vec{w}_3)^{\perp}, \\
(\vec{v}_0,\vec{v}_3)^{\perp} \otimes (\vec{w}_1,\vec{w}_2,\vec{w}_4)^{\perp}, \\
(\vec{v}_0, \vec{v}_2)^{\perp} \otimes (\vec{w}_1,\vec{w}_3,\vec{w}_4)^{\perp}.
\ea
\ee
These four vectors are not enough to span the full Hilbert space
${\cal H}_{{\bf Pyr34}^+}^{\perp}$. This means that the range of $\bar{\rho}_{{\bf Pyr34}^+}$ contains
only four product states, whereas $\bar{\rho}_{{\bf Pyr34}^+}$ has rank 6. Therefore $\bar{\rho}_{{\bf Pyr34}^+}$ must be entangled. The entanglement of
$\bar{\rho}_{{\bf Pyr34}^+}$ is bound by construction. Since $\bar{\rho}_{{\bf Pyr34}^+}$ is entangled, Proposition \ref{cosep} implies that the set {\bf Pyr34}$^+$ is a SUCPB.
 
So we have constructed a new bound entangled state whose range is
not exempt from product states but has a product state {\em deficit}.
This set is the first example of a bound entangled state related to a SUCPB, which is not a UPB. {\bf Pyr34}$^+$ shares with any UPB the fact that its members cannot be
distinguished perfectly by local POVMs and classical communication.
In conclusion, we have gone from a UCPB {\bf Pyr34} to a SUCPB {\bf Pyr34}$^+$, or from a separable
state $\bar{\rho}_{{\bf Pyr34}}$ to a bound entangled state  $\bar{\rho}_{{\bf Pyr34}^+}$. This construction is an example of a general way to make a
bound entangled state from a UCPB:
 
\begin{lem}
Given a UCPB {\rm S} on ${\cal H}=\bigotimes_{i=1}^m {\cal H}_i$ there always exists a (possibly empty) set of mutually orthogonal product states orthogonal to {\rm S} such that when added to {\rm S} to make ${\rm S}^+$, the density matrix $\bar{\rho}_{{\rm S}^+}$ is bound entangled.
\label{splus}
\end{lem}
 
\noindent {\it Proof}. We consider the density matrix $\bar{\rho}_{\rm S}$ which is either
separable or bound entangled. If it is separable then there exists
at least one product state in the range of $\bar{\rho}_{\rm S}$. We add
this state to S and repeat this procedure until the projector onto the
complementary subspace of this augmented set is entangled.  When S is uncompletable, then we cannot keep adding orthogonal
product states: If we would be able to add orthogonal product states
until we have a full product basis for ${\cal H}$, then the set S would
be completable on the given Hilbert space ${\cal H}$, which is in
contradiction with S being a UCPB. $\Box$
 
The lemma leaves open the possibility that the only
bound entangled density matrices $\bar{\rho}_{{\rm S}^+}$ we can find are
when S has been extended all the way into a UPB. Our example {\bf Pyr34}$^+$
shows that this is not always the case.
 
One question of interest which we have not been able to answer is the
following: Say we have a PB S which is a SUCPB, but not a UPB, such as
the set {\bf Pyr34}$^+$.  Will it be necessary to add more product states
to this set as Lemma \ref{splus} suggests to make a bound entangled
state on the complementary subspace? Or is the state $\bar{\rho}_{\rm S}$
where S is a SUCPB, but not a UPB, {\em always} bound entangled? 

\begin{figure}[htbf]
\centering
\epsfxsize=3.5in
\leavevmode\epsfbox{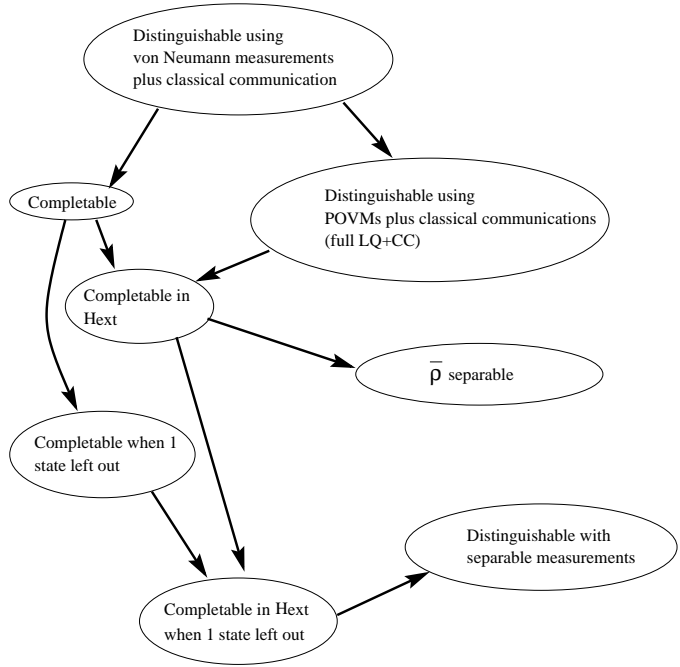}
\caption{The network of positive implications of the results discussed in
section \protect\ref{properties} and \protect\ref{usesepsup}.}
\label{implications}
\end{figure}

\begin{figure}[htbf]
\centering
\epsfxsize=4.0in
\leavevmode\epsfbox{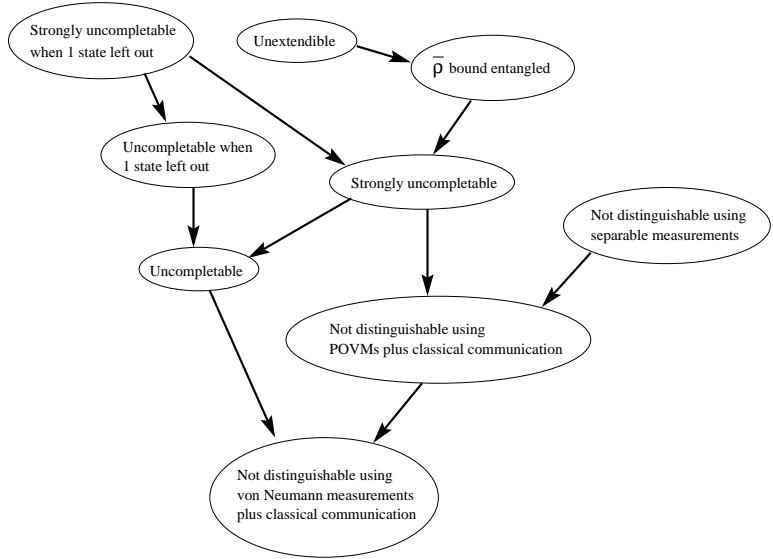}
\caption{The network of negative implications of the results discussed
in section \protect\ref{properties} and \protect\ref{usesepsup}.}
\label{implications2}
\end{figure}

 
In the Figures \ref{implications} and \ref{implications2} we show the
network of relations that was partially discussed in this section.
In section \ref{usesepsup} we will discuss one of these relations,
which is the question when orthogonal product states are distinguishable
by separable superoperators.
 
\section{The Use of Separable Superoperators}
\label{usesepsup}
 
In this section we address the question of what kind of measurement
{\em does} distinguish the members of a PB. We are interested in finding
measurements that need the least amount of resources in terms of
entanglement between the two or more parties. We introduce a class
of quantum operations that are close relatives of operations that can
be implemented by local quantum operations and classical
communication, the separable superoperators and measurements:

\begin{defi}\cite{rainssep}
Let ${\cal H}=\bigotimes_{i=1}^n {\cal H}_i$. Let ${\cal H}'=\bigotimes_{i=1}^n {\cal H}'_i$. A ${\rm TCP}$ map ${\cal S}\colon\,B({\cal H})\rightarrow B({\cal H}')$ is separable if and only if one can write the
action of ${\cal S}$ on any density matrix $\rho \in B({\cal H})$ as
\be
{\cal S}(\rho)=\sum_i A_{1,i} \otimes A_{2,i} \otimes \ldots \otimes A_{n,i}\, \rho
\,A_{1,i}^{\dagger} \otimes A_{2,i}^{\dagger} \otimes \ldots \otimes A_{n,i}^{\dagger},
\label{sepsup}
\ee
where the ``operation element'' $A_{k,i}$ is a ${\rm dim}{\cal H}'_i \times {\rm dim}{\cal H}_i$ matrix
and
\be
\sum_i  A_{1,i}^{\dagger} A_{1,i} \otimes  A_{2,i}^{\dagger} A_{2,i} \otimes \ldots \otimes A_{n,i}^{\dagger} A_{n,i}={\bf 1}.
\ee
Similarly, a quantum measurement on a
multipartite Hilbert space is separable if and only if for each outcome $m$, the
operation elements $A_i^m$ for all $i$ are of a separable form:
\be
A_i^m=A_{1,i}^m \otimes A_{2,i}^m \otimes \ldots \otimes A_{n,i}^m.
\ee
\end{defi}
 
Testing whether or not a superoperator is separable is not a simple problem
since the operation elements $A_i$ of a superoperator ${\cal S}$ are not uniquely defined.
The results of Ref. \cite{qne} show that separable superoperators are not
equivalent to local quantum operations and classical communication.
There is a separable measurement for the nine states presented in Ref. \cite{qne};
it is the measurement whose operation elements are the projectors onto
the nine states. But the nine states are not locally distinguishable by LQ+CC.
 
The following theorem gives a sufficient condition under which a set
of bipartite orthogonal product states is distinguishable with the use
of separable measurements.  Unfortunately, it is not known what
entanglement resources are needed to implement such separable
measurements. They do however form a rather restricted class of
operations. Since they map product states onto product states it is
not possible to use them to create entanglement where none previously existed.

\begin{theo}
Let {\rm S} be a bipartite PB in
${\cal H}={\cal H}_A \otimes {\cal H}_B$ with $k$ members. If {\rm S} has the property that it is completable
in ${\cal H}$ or local extensions of ${\cal H}$ (${\cal H}_{ext}$) when
any single member is removed from {\rm S}, then the members of {\rm S} are
distinguishable by means of a separable measurement.
\label{povmmeasu}
\end{theo}

\noindent {\it Proof}. Denote the orthogonal rank 1 product projectors
onto the states in S as $\{\Pi_m\}_{m=1}^{k}$. Let S$_i$, $i=1,\ldots,k$ be 
the set S without a particular state $i$.
Since each set S$_i$ is completable, the (unnormalized) states
\be
\Pi_{{\rm S}_i^{\perp}}={\bf 1}-\sum_{k \neq i}\Pi_k
\ee
for $i=1,\ldots,k$ are separable. Note that $\Pi_{{\rm S}_i^{\perp}}=\Pi_{{\rm S}_i^{\perp}}^{\dagger}$. The projectors $\Pi_{{\rm S}_i^{\perp}}$ and $\Pi_i$ for
$i=1,\ldots,k$ can be made to sum up to the identity by choosing the right coefficients:
\be
\frac{1}{k} \sum_{i=1}^k \Pi_{{\rm S}_i^{\perp}}^{\dagger} \Pi_{{\rm S}_i^{\perp}} + \frac{k-1}{k} \sum_{i=1}^k \Pi_i^{\dagger}\Pi_i ={\bf 1},
\label{sumto1}
\ee
using $\Pi^2=\Pi$ for projectors.
Since the projectors $\Pi_{{\rm S}_i^{\perp}}$ are separable, one can decompose them
into a set of $N_i$ rank 1 product projectors,
$\Pi_{({\rm S}_i^{\perp}, m_i)}$ labeled by an index $m_i=1,\ldots,N_i$.  Note that one can choose mutually orthogonal projectors (for a given $i$)
$\Pi_{({\rm S}_i^{\perp},m_i)}$ when ${\rm S}_i$ is completable in the given Hilbert
space ${\cal H}$. When ${\rm S}_i$ is completable only in a local extension of
${\cal H}$, these projectors will be non-orthogonal. In both cases the
set of product projectors
\be
\left\{\frac{1}{\sqrt{k}}\Pi_{({\rm S}_i^{\perp},m_i)}\,,\,\sqrt{\frac{k-1}{k}} \Pi_i\right\}_{i=1,m_i=1}^{k,N_i},
\label{meassepsup}
\ee
are the operation elements of a separable measurement. This measurement 
pro\-jects onto states in S or onto product states that are orthogonal to all but one state in S. With a slight modification of this measurement one can
construct a measurement which distinguishes the states in S locally. Formally
one replaces the projectors of Eq. (\ref{meassepsup}) by
\be
\ba{c}
\Pi_i=\ket{\alpha_i,\beta_i}\bra{\alpha_i,\beta_i} \rightarrow \ket{i_A,i_B}\bra{\alpha_i,\beta_i},\\
\Pi_{(S_i^{\perp},m_i)}=\ket{\delta_{i,m_i},\gamma_{i,m_i}}\bra{\delta_{i,m_i},\gamma_{i,m_i}} \rightarrow
\ket{i',{m_i}_A,i',{m_i}_B}\bra{\delta_{i,m_i},\gamma_{i,m_i}},
\ea
\ee
such that the set of states $\ket{i_A}$, $\ket{i',m_{i_A}}$ is an orthonormal
set for A and the same for B. This modification leaves Eq. (\ref{sumto1})
unchanged, so that this new set of operation elements again corresponds to a
(separable) measurement. Upon this measurement, however, Alice and Bob both
get a classical record of the outcome. If they perform this measurement on
states in S, their outcomes will uniquely determine which state in S they
were given. $\Box$
 
We will show in section \ref{sixparam}, using the method of orthogonality graphs, that all UPBs in $3 \otimes 3$ have exactly five members.
Theorem \ref{no34} (see also section \ref{orthograph}) tells us that any
set of four orthogonal bipartite product states is completable. Therefore
Theorem \ref{povmmeasu} implies that all UPBs in $3 \otimes 3$ are distinguishable by a separable measurement.

\section{The Orthogonality Graph of a Product Basis}
\label{orthograph}
 
It is convenient to describe the orthogonality structure of a
set of orthogonal product states on a multipartite Hilbert space
by a graph, which we will call the {\em orthogonality graph} of the PB:
Essentially the same graph has appeared in a connection with a problem 
in classical information theory \cite{lovasz}.

\begin{defi}
Let ${\cal H}=\bigotimes_{i=1}^m {\cal H}_i$ be a $m$-partite Hilbert space
with $\dim {\cal H}_i=d_i$. Let $S=\{(\psi_j\!\equiv\!\bigotimes_{i=1}^m\varphi_{i,j}):j=1\ldots n\}$ be an orthogonal product basis (PB) in ${\cal H}$.
We represent ${\rm S}$ as a graph $G=(V,E_1 \cup E_2 \cup \ldots \cup
E_m)$ where the set of edges $E_i$ have color $i$. The states $\psi_j \in {\rm S}$
are represented as the vertices $V$. There exists an edge $e$ of color $i$
between the vertices $v_k$ and $v_l$, i.e. $e \in E_i$,  when states $\psi_k$ and
$\psi_l$ are orthogonal on ${\cal H}_i$. Since all the states in the PB are
mutually orthogonal, every vertex is connected to all the other vertices
by at least one edge of some color.
\end{defi}
 
\begin{figure}[htbf]
\centering
\epsfxsize=3.5in
\leavevmode\epsfbox{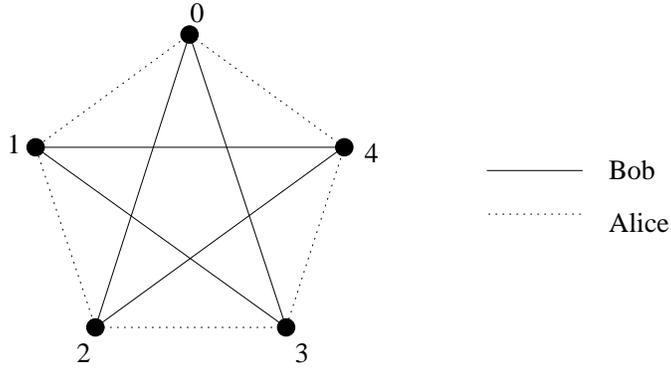}
\caption{The orthogonality graph of any UPB on $3 \otimes 3$.}
\label{pentfig}
\end{figure}

An example of an orthogonality graph is given in Fig. \ref{pentfig}; it is
the graph for the bipartite {\bf Pyramid} UPB. Note that it is also possible
to have several edges of different colors between two vertices when
states are orthogonal for more than one party.
 
The representation of a PB in terms of a graph can be useful when
we want to determine whether the members of the PB are distinguishable by
means of local operations and classical communication.
By enumerating the possible orthogonality graphs, it is not hard to prove
the following
 
\begin{propo}
The members of any multipartite PB {\rm S} with three or
fewer members are distinguishable by local incomplete von Neumann
measurements and classical communication, and the PB is completable to
a full product basis.
\label{oneprime}
\end{propo}
 
\begin{figure}[htbf]
\centering
\epsfxsize=4in
\leavevmode\epsfbox{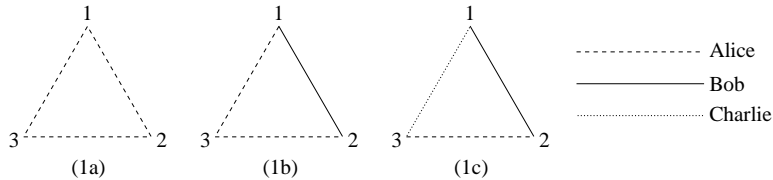}
\caption{The possible orthogonality graphs of a multipartite PB with three
members.}
\label{graphsmulti}
\end{figure}

\noindent {\it Proof}. We first note that we need only show that the
members of S are distinguishable by local von Neumann measurements to
also show that S is completable because of Lemma \ref{immeasurability}.
Now, when S has one member, there is nothing to distinguish and the
statement is trivial.  With two members, the states must be orthogonal
for some party and that party can distinguish them.  With three
members the possible orthogonality graphs are depicted in
Fig. \ref{graphsmulti}.  We have omitted graphs with multiply colored
edges.  A multiply colored edge can only make it easier to distinguish
the members of the corresponding PB. Thus when a graph corresponds to
a distinguishable set after we leave out any multiple coloring, it also
corresponds to a distinguishable set with the multiple coloring.  We
have similarly omitted graphs which are the same as the graphs shown
under interchange of parties as clearly those cases will follow the
same line of reasoning.

In case (1a), as all the states are mutually
orthogonal on Alice's side, Alice can do a measurement that uniquely
distinguishes them. In case (1b) the third state is orthogonal to
both state 1 and state 2 on Alice's side. Therefore Alice can distinguish
between $(1,2)$ and $3$. Then Bob can finish the measurement by
telling apart 1 and 2 locally.  In case (1c) Alice distinguishes
state 2 from state 3, Bob distinguishes 1 from 2, and Charlie
distinguishes 1 from 3, together determining the state. $\Box$
 
This proposition cannot be strengthened any
further: a three party UPB exists with only four members, it is the set
{\bf Shifts} (Eq. (\ref{defshifts}) and Fig. \ref{graphs}(a)). However, a
stronger result may be obtained in the case of a {\em bipartite} PB:
 
\begin{theo}
Let ${\rm S}$ be a bipartite PB with four or fewer members, i.e.
$|{\rm S}| \leq 4$ in any dimension (that allows for this PB).
The set {\rm S} is distinguishable by local incomplete
von Neumann measurements and classical communication. The set ${\rm S}$ is
completable to a full product basis for ${\cal H}$.
\label{no34}
\end{theo}
 
\begin{figure}[htbf]
\centering
\epsfxsize=4in
\leavevmode\epsfbox{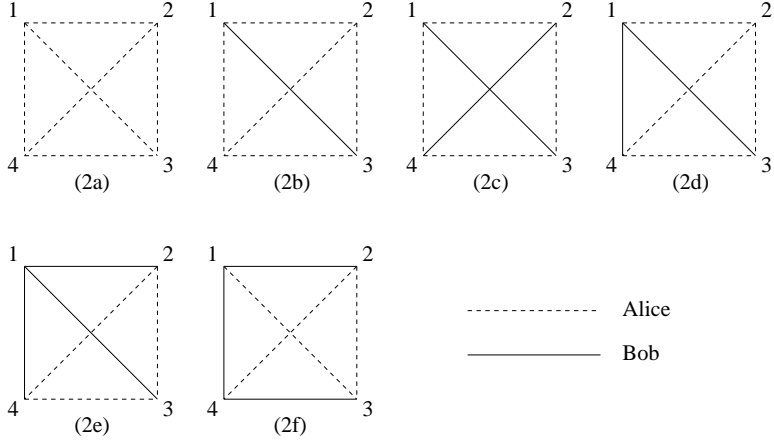}
\caption{The possible orthogonality graphs of a bipartite PB with four members.}
\label{3mem}
\end{figure}

\noindent {\it Proof}. We will expand on the proof of Proposition
\ref{oneprime}.  When the set S has one, two, or three members,
Proposition \ref{oneprime} applies directly. When S has four members the
six possible orthogonality graphs are as given in
Fig. \ref{3mem}. Again we omit graphs that are identical to these six
under interchanging of parties, and graphs with doubly colored edges.
 
Case (2a) is trivial.  In cases (2b), (2d), and (2e) there is always a state
that is orthogonal to all the other states on one side.  The measurer
associated with that side can then distinguish this state from all the
others.  The result is that three states are left to be distinguished, which is
covered by Proposition \ref{oneprime}.
 
In case (2c) $(1,3)$ can be distinguished from $(2,4)$ on Alice's side
after which we are left with two orthogonal states that can be locally
distinguished by Bob. In Case (2f) a different type of measurement must
be carried out. In the previous cases the measurements
were such that none of the states were changed by the measurement. The
set of states S was simply {\em dissected} in subsets. However, Alice and Bob
can carry out a more general type of measurement, namely
one that can {\em change} the states. Such a measurement must be
orthogonality-preserving; by this we mean that the changed states
that are left over to be distinguished in a succeeding round must remain
orthogonal under the measurement.
In case (2f) state 2 is orthogonal to both states 3
and 4 on Alice's side. Let Alice project with $\Pi_{34}$, the
projector on the subspace spanned by her side of states 3 and 4, and
with $\Pi_2$, the projector on her side of state 2, and possibly
$\Pi_{else}$ where $\Pi_{else} \Pi_{34}=0$, $\Pi_{else} \Pi_{2}=0$ and
$\Pi_{else}+\Pi_{34}+\Pi_2={\bf 1}$. The projector $\Pi_{else}$ is only used when
the states 2, 3 and 4 do not yet span the full Hilbert space of Alice;
if this outcome is obtained, state 1 has been conclusively identified.
Otherwise,
state 1 is mapped onto $\Pi_{34} \ket{1}$ or
$\Pi_{2}\ket{1}$. If the outcome is 2, Bob can finish the protocol by
locally distinguishing 1 and 2.  If the outcome is 34 we notice that
we are in a three state case again {\em and} all states are still
mutually orthogonal; $\Pi_{34}\ket{1}$ and state 3 are still
orthogonal on Alice's side as state 3 is invariant under this
projection $\Pi_{34} \ket{3}=\ket{3}$. $\Box$
 
These preliminary results will now be used to give a complete
characterization of UPBs in $3 \otimes 3$.
 
\subsection{A Six-parameter Family of UPBs in $3 \otimes 3$}
\label{sixparam}
 
\begin{figure}[htbf]
\centering
\epsfxsize=2.5in
\leavevmode\epsfbox{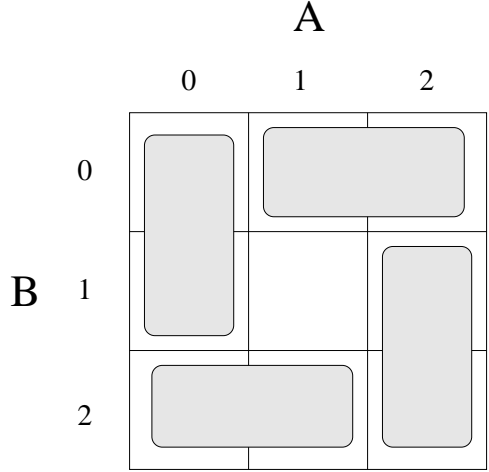}
\caption{Tile structure of the bipartite $3 \otimes 3$ \protect {\bf Tiles} UPB.}
\label{Tiles}
\end{figure}

In Ref. \cite{upb1} we presented two examples of  UPBs in $3 \otimes 3$.
One is the {\bf Pyramid} set which was discussed in section \ref{defin} and
the second was the set {\bf Tiles}. The following five states on $3 \otimes 3$ form a UPB denoted as {\bf Tiles}
\be
\ba{lr}
\ket{\psi_0}={1 \over
\sqrt{2}}\ket{0}(\ket{0}-\ket{1}),&\ \ \ket{\psi_2}= {1 \over
\sqrt{2}}\ket{2}(\ket{1}-\ket{2}),\\ \ket{\psi_1}={1 \over
\sqrt{2}}(\ket{0}-\ket{1})\ket{2},&\ \ \ket{\psi_3}= {1 \over
\sqrt{2}}(\ket{1}-\ket{2})\ket{0},\\ \ \ \ \
\lefteqn{\ket{\psi_4}=(1/3)(\ket{0}+\ket{1}+\ket{2})(\ket{0}+\ket{1}+\ket{2}).}
\ea
\ee
Note that the first four states are the interlocking tiles of
Ref. \cite{qne}, and the fifth state works as a ``stopper'' to force the
unextendibility. The set can be represented
with the use of tiles as in Fig. \ref{Tiles}. A tile can
represent one or more states. For example, the tile in the upper left
corner of Fig. \ref{Tiles} represents a state which is of the form
\be
\ket{0} \otimes (\alpha_0\ket{0}+\alpha_1 \ket{1}).
\ee
The ``stopper'' state is not included in the figure; as a tile it would
cover the full square.
 
These two sets, {\bf Pyramid} and {\bf Tiles}, are examples of a larger six-parameter family of unextendible product bases in $3 \otimes 3$. We will
prove that this six-parameter family gives an exhaustive characterization
of UPBs in $3 \otimes 3$. First we note that five is the smallest number
of states in a UPB in $3 \otimes 3$, due to Eq. (\ref{min}).
 
We will now prove that any UPB with five members on $3 \otimes 3$
must have an orthogonality graph as in Fig. \ref{pentfig}.
We will do so by arguing that any vertex must be connected to exactly
two other vertices by an edge of the same color. The argument goes as follows.
If there exists a vertex that is connected to four other vertices with edges of a single color, then we can locally distinguish this state from the other
four states. Theorem \ref{no34} implies that we can also distinguish the
remaining four states. Now, assume that there
exists a vertex, say vertex 1, that is connected to three other vertices,
corresponding, say, to the states 2, 3 and 4. Then we can distinguish between
1 and $(2,3,4)$ by a local projection that splits state 5 in two projected
states. However this projected state 5 is still orthogonal to 1 and $(2,3,4)$. Thus we are left with distinguishing a set of two or four orthogonal product states which
can be done locally by Theorem \ref{no34}.
 
Finally, it is not hard to see that if all vertices have to be connected to exactly two
other vertices, the orthogonality
graph in Fig. \ref{pentfig} is the only possible graph.
 
Now that we have established a unique orthogonality graph, it remains to characterize the solution set. Let $\ket{\psi_i}=\ket{\alpha_i}\otimes \ket{\beta_i}$, $i=0,\ldots, 4$.
Let $(\gamma_A,\theta_A,\phi_A,\gamma_B,\theta_B,\phi_B)$ be a set of
six angles. We set
\begin{eqnarray}
&&\ket{\alpha_0}=\ket{0},\nonumber\\
&&\ket{\alpha_1}=\ket{1},\nonumber\\
&&\ket{\alpha_2}=\cos\theta_A\ket{0}+\sin\theta_A\ket{2},\nonumber\\
&&\ket{\alpha_3}=\sin\gamma_A\sin\theta_A\ket{0}-
\sin\gamma_A\cos\theta_A\ket{2}+\cos\gamma_Ae^{i\phi_A}\ket{1},\nonumber\\
&&\ket{\alpha_4}={1\over N_A}(\sin\gamma_A\cos\theta_Ae^{i\phi_A}\ket{1}+
\cos\gamma_A\ket{2}),\nonumber\\
&&\ket{\beta_0}=\ket{1},\nonumber\\
&&\ket{\beta_1}=\sin\gamma_B\sin\theta_B\ket{0}-
\sin\gamma_B\cos\theta_B\ket{2}+\cos\gamma_Be^{i\phi_B}\ket{1},\nonumber\\
&&\ket{\beta_2}=\ket{0},\nonumber\\
&&\ket{\beta_3}=\cos\theta_B\ket{0}+\sin\theta_B\ket{2},\nonumber\\
&&\ket{\beta_4}={1\over N_B}(\sin\gamma_B\cos\theta_Be^{i\phi_B}\ket{1}+
\cos\gamma_B\ket{2}),\label{defs}
\end{eqnarray}
with normalizations
\be
N_{A,B}=\sqrt{\cos^2\gamma_{A,B}+\sin^2\gamma_{A,B}\cos^2\theta_{A,B}}.
\ee
We have taken
$\ket{\alpha_{0,1}}$ to define the first two vectors $\ket{0}, \ket{1}$ of
the Alice Hilbert space; the overall phase of $\ket{\alpha_2}$ and $\ket{\alpha_3}$ and the
phase of the $\ket{2}$ vector are chosen so that $\ket{\alpha_2}$, and the first two terms
of the above expression for $\ket{\alpha_3}$, are real.  Also, the
overall phase of $\ket{\alpha_4}$ is fixed so that the coefficient of
$\ket{2}$ is real.  All the same remarks apply correspondingly to
the Bob states. In order for this set of states to be a UPB we require
that $\cos \theta_{A,B} \neq 0$,  $\cos \gamma_{A,B} \neq 0$,
$\sin \theta_{A,B} \neq 0$ and $\sin \gamma_{A,B} \neq 0$. If this
restriction is made, we see that any set of three different vectors for Alice
or for Bob spans a three dimensional space. The {\bf Pyramid} UPB is obtained from Eq. (\ref{defs}) with the
parameter choices $\phi_{A,B}=0$, $\theta_{A,B}=
\gamma_{A,B}=\cos^{-1}((\sqrt{5}-1)/2)$.  The parameters for the {\bf Tiles} UPB are $\phi_{A,B}=0$, $\theta_{A,B}=\gamma_{A,B}=3\pi/4$.
 
We find that all the solutions having the orthogonality graph of Fig. \ref{pentfig}
correspond to UPBs. If we had lifted the restriction on the angles,
say, setting $\sin \theta_A=0$, then the set would no longer be a UPB as
$\ket{\alpha_2} \in {\rm span}(\ket{\alpha_0},\ket{\alpha_1})$. At the same
time the set would no longer correspond to the graph of Fig. \ref{pentfig},
as now state $\ket{\alpha_2}$ is orthogonal to $\ket{\alpha_4}$.
 
This suggests that UPBs can be characterized by their orthogonality graphs;
when a set of states S has an orthogonality graph $G$ and S is a
UPB then all the sets with graph $G$ are UPBs. If this conjecture were true, it
would imply that we can classify UPBs by their orthogonality graphs leading
to an important simplification. But in section \ref{septetc} we present a
counterexample to this conjecture for three parties and seven states.
 
Finally to finish the characterization, we prove that any PB with six or more members in $3 \otimes 3$ is completable. We give the proof excluding a
six member UPB in Appendix \ref{nosixmem}. The density matrix $\bar{\rho}_{PB}$ of a PB with seven or eight states has rank two and rank one respectively. By construction this density matrix is either a bound entangled state or a separable state, as follows
from Theorem~\ref{rhobar}. It can be shown by different arguments that there exists no bound entangled state with rank less than or equal to two \cite{norank2}.
The state must therefore be separable. To the seven state PB we therefore
can add a product vector to make it an eight state PB which is again extendible.

\section{Multipartite and high dimensional bipartite UPBs}
\label{manyexam}
 
In this section we introduce several examples and families of UPBs. In
section \ref{genshifts} we present UPBs on multi-qubit Hilbert spaces.
In section \ref{gentiles} we give two constructions of high dimensional
bipartite UPBs based on tiling patterns such as in Fig. \ref{Tiles}.
In section \ref{septetc} we give multipartite UPBs based on a
generalization of the orthogonality graph of the {\bf Pyramid} UPB
(Fig. \ref{pentfig}).  In section \ref{quadresi} we present
a bipartite high dimensional UPB which is based on quadratic residues.
Finally, in section \ref{tpupbsec} we prove that tensor products of
UPBs are again UPBs.
 
\subsection{{\bf GenShifts} and other UPBs in qubit Hilbert spaces}
\label{genshifts}
 
We first give a theorem that was proved in Ref. \cite{upb1}:
 
\begin{theo}\cite{upb1} Any set of orthogonal product states $\{\ket{\alpha_i} \otimes \ket{\beta_i}\}_{i=1}^k$ in $2 \otimes n$ for any $n \geq 2$ is distinguishable by local measurements and classical communication and therefore completable to a full product basis for $2 \otimes n$.
\label{nogo2n}
\end{theo}
 
Even though any bipartite product basis involving a qubit Hilbert space
is completable, we found in Ref. \cite{upb1} that a tripartite UPB involving
three qubits is possible. This was the set {\bf Shifts} given by the states
\be
\{\ket{0,0,0}, \ket{+,1,-}, \ket{1,-,+}, \ket{-,+,1}\}.
\label{defshifts}
\ee
 
It follows from Theorem \ref{nogo2n} that when we make {\em any} bipartite
split of the three parties, say we join parties BC, that the set {\bf Shifts}
is completable to a full product basis for ${\cal H}_A \otimes {\cal H}_{BC}$.
Thus the bound entangled state that we construct from {\bf Shifts} as in
Eq. (\ref{compl}), must be separable over any bipartite split.
Therefore this bound entangled state could have been made without any
entanglement between A and BC, {\em or} AB and C, {\em or} C and AB. However, the
state is entangled. One may say that the entanglement is delocalized
over the three parties.
 
Our construction of {\bf Shifts} can be generalized to multipartite UPBs,
which we will call {\bf GenShifts}. Again, because of Theorem \ref{nogo2n}, the
bound entangled states based on {\bf GenShifts} have a form of delocalized
entanglement. The bound entangled states could have been made without
entanglement between any single party and all the other remaining parties.
We do not know whether the bound entangled states are
separable over a split in two or more parties and all the other parties.
 
{\bf GenShifts} is a \oupb on $\bigotimes_{i=1}^{2k-1} {\cal H}_2$ with $2k$
members, the minimal number for a UPB, Eq. (\ref{min}). The first state is $\ket{0,\ldots 0,0}$. The second is
\be
\ket{1, \psi_1, \psi_2, \ldots, \psi_{k-1}, \psi_{k-1}^{\perp}, \ldots
,\psi_2^{\perp}, \psi_1^{\perp}}.
\ee
The states $\ket{\psi_i}$ and $\ket{\psi_j}$ for all $i \neq j $ are chosen to be neither
orthogonal nor identical. Also, $\ket{\psi_i}$ is neither orthogonal nor
identical to the state $\ket{0}$ for all $i$. The other states in
the \oupb are obtained by (cyclic) right shifting the second state, i.e. the
third state is
\be
\ket{\psi_1^{\perp}, 1, \psi_1, \psi_2, \ldots,\psi_{k-1}, \psi_{k-1}^{\perp}, \ldots,\psi_2^{\perp}}.
\ee
These states are all orthogonal in the following way: The state
$\ket{0, \ldots, 0,0}$ is special and it is orthogonal to all the
other states as they all have a $\ket{1}$ for some party. Leaving this
special state aside, all states are orthogonal to the next state,
their first right-shifted state, by the orthogonality of
$\ket{\psi_{k-1}}$ and $\ket{\psi_{k-1}^{\perp}}$. All states are
orthogonal to the $2$nd right-shifted state by the orthogonality of
$\ket{\psi_1}$ and $\ket{\psi_1^{\perp}}$. The $3$rd right-shifted
state is made orthogonal with $\ket{\psi_{k-2}}$ and
$\ket{\psi_{k-2}^{\perp}}$. We can continue this until the last
$(2k-2)$th right-shifted state and we are done.
 
As there are no states repeated on one side of the \oupb all sets of two
states span a two dimensional space; Lemma \ref{rule1} implies that
the set is a UPB.
 
\begin{figure}[htbf]
\centering
\epsfxsize=4.0in
\leavevmode\epsfbox{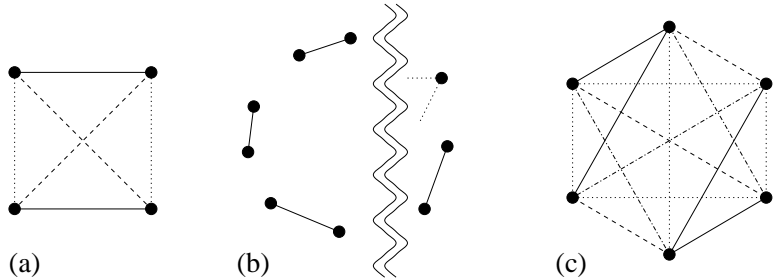}
\caption{Orthogonality graphs for qubit UPBs.  (a) {\bf Shifts}, i.e. {\bf GenShifts} for $k=2$.
(b) Demonstration of the nonexistence of a minimal UPB with an
even number of qubits.  (c) A six-state, 4-partite UPB.}
\label{graphs}
\end{figure}

The orthogonality graph for the first example of {\bf GenShifts} which is
just the set {\bf Shifts}, is shown in
Fig. \ref{graphs}(a).  For $k=2$ and $k=3$, the graph for {\bf GenShifts} is
the only orthogonality graph possible for a UPB in this Hilbert space.  For
$k>3$ graphs other than the one corresponding to {\bf GenShifts} are
possible.  It is simple to argue that, as in the $3\otimes 3$ UPB (section \ref{sixparam}), all PBs having the {\bf GenShifts} orthogonality graph are UPBs:
The orthogonality graph of {\bf GenShifts} is partially characterized by the
fact that there is only one edge emanating from every vertex of a particular
color. This implies that no states in a set corresponding to this orthogonality graph are repeated, that is, the same state for a party $i$ is not used more than once in
the set. As the set is minimal, this implies that such an orthogonality
graph directly fulfills the conditions of Lemma \ref{rule1}; every pair of two
states spans a two dimensional space. Also, when we consider a
minimal PB (having $n+1$ members for $n$ parties) and its orthogonality
graph has a doubly colored edge, then the PB cannot be a UPB. This is because
the property of having a doubly colored edge directly translates into
some pair of states not spanning a two dimensional Hilbert space.
 
Using the orthogonality-graph construction, we can prove that if the
number of parties $n$ is even, then qubit UPBs with $n+1$ states, the
minimal possible number by Lemma 1, do not exist.  We show this by
demonstrating that some states would have to be repeated; but repeated
states permit a partitioning as in Lemma 1 which allows the
introduction of another orthogonal product state.
Fig. \ref{graphs}(b) illustrates the idea: Considering the lines of
just one color, we note that two cannot emanate from the same node
(otherwise there would be a repeat), but after joining them up
pairwise there will be one left over, since the number of states is
odd.  But since this last node has no lines of the first color coming
into it, it will have to have at least two of some other color
emerging from it, which would again force a repeat.  Therefore, the
basis would be extendible.
 
On the other hand, non-minimal UPBs for even numbers of qubits do
exist; Fig. \ref{graphs}(c) shows the graph for one with six states in a space of four qubits. See Ref. \cite{lovalon} for results on the existence of 
minimal UPBs in multipartite Hilbert spaces of arbitrary dimension. 

\subsection{{\bf GenTiles}}
\label{gentiles}
 
\begin{figure}[htbf]
\centering
\epsfxsize=3.5in
\leavevmode\epsfbox{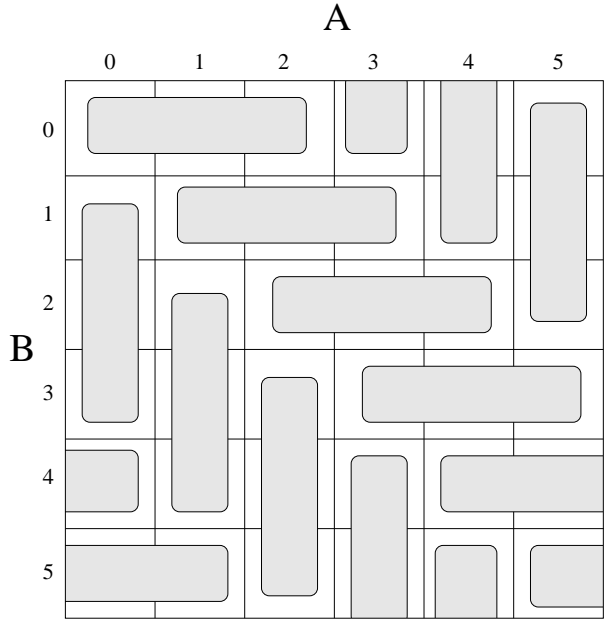}
\caption{Tile structure of the bipartite $6 \otimes 6$ UPB.}
\label{tile}
\end{figure}

We introduce a bipartite product basis {\bf GenTiles1} in $n \otimes n$
where $n$ is even. These states have a tile structure which in the
case of $6 \otimes 6$ is shown in Fig. \ref{tile}. The general construction goes as follows: We label a set of $n$
orthonormal states as $\ket{0},\ldots, \ket{n-1}$.  We define the set
of `vertical tile' states
\bea
\ket{V_{mk}}=\ket{k}\otimes \ket{\omega_{m,k+1}}=\ket{k} \otimes
\sum_{j=0}^{n/2-1} \omega^{jm} \ket{j+k+1 \bmod n}, \nonumber \\
m=1,\ldots ,n/2-1, \;\;k=0,\ldots,n-1,
\eea
where $\omega=e^{i 4 \pi/n}$. Similarly, we define the set of `horizontal tile' states:
\be
\ket{H_{mk}}=\ket{\omega_{m,k}} \otimes \ket{k},\;\;\; m=1,\ldots ,n/2-1,\;\;k=0,\ldots,n-1.
\ee
Finally we add a `stopper' state
\be
\ket{F}=\sum_{i=0}^{n-1} \sum_{j=0}^{n-1}\ket{i} \otimes \ket{j}.
\ee
The stopper state is not depicted in Fig. \ref{tile}; as a tile it
would cover the whole $6$ by $6$ square. The representation of the set
as an arrangement of tiles informs us about the orthogonalities among
some of its members. It is not hard to see that nonoverlapping tiles
are orthogonal.  The orthogonality of the states $\ket{V_{mk}}$ and
$\ket{V_{m'k}}$ for $m \neq m'$ follows from the identity
\be
\langle \omega_{m,k}\,|\,\omega_{m',k} \rangle \propto
\delta_{mn}.
\ee
With the same identity we can prove that the states  $\ket{H_{mk}}$ and  $\ket{H_{m'k}}$ for $m \neq m'$ are mutually orthogonal. Finally, every state
$\ket{H_{mk}}$ or $\ket{V_{mk}}$ is orthogonal to the `stopper' $\ket{F}$ as
\be
\sum_{j=0}^{n/2-1} \omega^{jm} \propto \delta_{m0},
\ee
and $m \neq 0$. The set has $n^2-2n+1$ states, much more than
the minimum number in a UPB on $n \otimes n$, which is $2n-1$. We can prove that this construction is a \oupb in $4 \otimes 4$ and $6 \otimes 6$ by
exhaustive checking of all partitions (see Lemma \ref{rule1}). This procedure
is hard to implement computationally for arbitrary high dimension, but one may conjecture (and prove, see Ref. \cite{terhaldiv:upbs}) that
 
\begin{theo}
The set of states {\bf GenTiles1} form a UPB on $n \otimes n$ for all even $n \geq 4$.
\end{theo}

\begin{figure}[htbf]
\centering
\epsfxsize=3.5in
\leavevmode\epsfbox{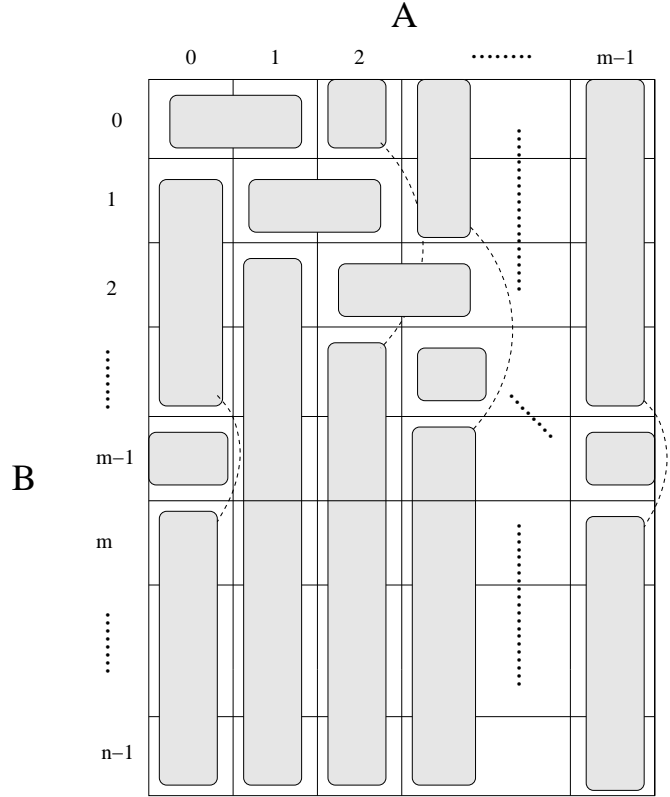}
\caption{Tile structure of the $m \otimes n$ {\bf GenTiles2} PB.}
\label{tilemn}
\end{figure}

Another tile construction which we call {\bf GenTiles2} can be made in dimensions $m \otimes n$ for $n > 3$, $m \geq 3$ and $n\geq m$.
The construction is illustrated in Fig. \ref{tilemn}. The small tiles which
cover two squares are given by
\be
\ket{S_j}=\frac{1}{\sqrt{2}} (\ket{j}-\ket{j+1 \bmod m}) \otimes \ket{j},\;\;\;0 \leq j \leq m-1.
\label{tileSdef}
\ee
These short tiles are mutually orthogonal on Bob's side.
The long tiles (in general not contiguous) that stretch out in the vertical
direction in Fig. \ref{tilemn} are given by
\bea
\ket{L_{jk}}=\ket{j} \otimes \frac{1}{\sqrt{n-2}}
\left(\sum_{i=0}^{m-3} \omega^{ik} \ket{i+j+1 \bmod m}+\sum_{i=m-2}^{n-3}\omega^{ik}\ket{i+2}\right), \nonumber \\
0 \leq j \leq m-1,\;\;1 \leq k \leq n-3,
\eea
with $\omega=e^{i \frac{2\pi}{n-2}}$. Lastly we add a `stopper' state
\be
\ket{F}=\frac{1}{\sqrt{nm}} \sum_{i=0}^{m-1}\sum_{j=0}^{n-1} \ket{i} \otimes \ket{j}.
\label{tileFdef}
\ee
The total number of states is $mn-2m+1$.

We can show that these states form a PB. For $j \neq j'$ the states $\ket{L_{j'k'}}$ and $\ket{L_{jk}}$ are orthogonal
on Alice's side. We also have
\be
\langle L_{jk'}\,|\,L_{jk} \rangle=\frac{1}{n-2} \sum_{p=0}^{n-3}e^{\frac{i 2\pi p(k-k')}{n-2}}=\delta_{kk'}.
\ee
The states $\ket{L_{jk}}$ and $\ket{S_{p}}$ with $p \neq j$ and $p \neq j+1 \bmod m$
are orthogonal on Alice's side. The long tiles $\ket{L_{jk}}$ are constructed such that they are orthogonal to the states $\ket{j}$ and $\ket{j+1 \bmod m}$ on
Bob's side, see Fig. \ref{tilemn}. Therefore $\ket{L_{jk}}$ is orthogonal to
$\ket{S_{j}}$ and $\ket{S_{j+1 \bmod m}}$. The states $\ket{S_l}$ are orthogonal to the
stopper $\ket{F}$ on Alice's side. Finally, the states $\ket{L_{jk}}$ are orthogonal
to $\ket{F}$ as
\be
\langle F\,|\,L_{jk} \rangle= \frac{1}{\sqrt{nm(n-2)}}\sum_{p=0}^{n-3}
e^{i \frac{2\pi pk}{n-2}}=\delta_{k 0},
\ee
and $k \neq 0$. We conjecture that this PB {\bf GenTiles2} is a UPB (the 
proof of the conjecture has been given in Ref. \cite{terhaldiv:upbs})
 
\begin{theo}
The set of states {\bf GenTiles2} form a UPB on $m \otimes n$ for
$n > 3$, $m \geq 3$ and $n \geq m$.
\end{theo}
 
Note that {\bf GenTiles2} with $m=n=3$ does {\em not} form a UPB.

We will now give some UPBs corresponding to generalizations of the
orthogonality graph in Fig. \ref{pentfig}. The first generalization is a UPB on $3 \otimes 3 \otimes \ldots \otimes 3$ (section \ref{septetc}).
The second generalization (section \ref{quadresi}) is another bipartite UPB in arbitrary high dimension.
 
\subsection{{\bf Sept} and {\bf GenPyramid}}
\label{septetc}
 
Let us first consider a generalization to $3 \otimes 3 \otimes 3$.
Define the following states
\be
\vec{v}_i=N (\cos{{ 2 \pi i \over 7}},\sin{{2 \pi i}\over 7},h),\;\;
i=0,\ldots,6,
\label{defsept}
\ee
with $h=\sqrt{-\cos \frac{4 \pi}{7}}$ and $N=1/\sqrt{1+|\cos \frac{4
\pi}{7}|}$.  The following seven states in $3 \otimes 3 \otimes 3$
form the \oupb {\bf Sept}
\be
\vec{p}_i = \vec{v}_i \otimes  \vec{v}_{2i\, \bmod\, 7}\otimes  \vec{v}_{3i\, \bmod\, 7},
\;\; i=0,\ldots,6.
\ee
 
\begin{figure}[htbf]
\centering
\epsfxsize=3.5in
\leavevmode\epsfbox{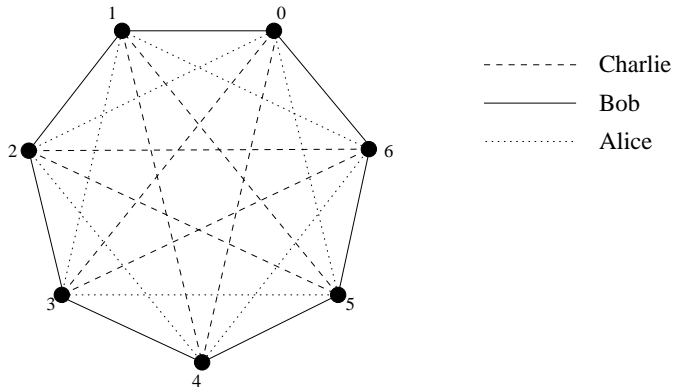}
\caption{The {\bf Sept} UPB on $3 \otimes 3 \otimes 3$.}
\label{sept}
\end{figure}

The orthogonality graph of these vectors $\vec{p_i}$ is shown in
Fig. \ref{sept}.  To prove that these states form a UPB, we must
show that any subset of three of them on one of the three sides (Lemma
\ref{rule1}) spans the full three dimensional space. As the vectors
$\vec{v}_i$ form the apex of a regular septagonal pyramid, there is no
subset of three of them that lies in a two dimensional plane.  It is
not known whether the complementary state $\bar{\rho}_{\bf Sept}$ is
separable over bipartite cuts, as with $\bar{\rho}_{\bf Shifts}$
(see Eq. (\ref{defshifts})), or whether it is a bound entangled over
the bipartite cuts.

This construction can be extended to $3^{\otimes n}$, the minimal UPB
thus constructed we will call {\bf GenPyramid}.  We have $n$ parties
and $p=2n+1$ states where $p$ is a prime number. Thus one can have
$(n,p)=(2,5),(3,7),(5,11)$, etc.
The states in the polygonal pyramid with $p$ vertices are defined as
\be
\vec{v}_i=N_p (\cos{{ 2 \pi i \over p}},\sin{{2 \pi i}\over p},
h_p),\;\;
i=0,\ldots,2n.
\label{defgensept}
\ee
In {\bf Sept} and {\bf Pyramid}, $h_p$ was chosen such that nonadjacent vertices
were orthogonal. For larger primes $p$ one has to make a choice
of which vectors to make orthogonal that depends
on $p$: in order for the vectors ${\vec v}_i$ and ${\vec v}_{i+m}$
to be made orthogonal by lifting these vectors out of
the plane of the polygon, we must have
\be
\frac{\pi}{2} \leq \frac{2 \pi m}{p} (\leq \pi),\label{boundm}
\ee
i.e. the angle between the vectors in the plane must be larger than
90 degrees. One can always find such an $m$ given a $p$, for example, for
$p=7$, $m=2$ or 3. With the choice of $m$ one fixes $h_p$ and $N_p$ as
\be
\ba{lr}
h_p=\sqrt{-\cos \frac{2 \pi m}{p}}, & N_p=1/\sqrt{1+|\cos\frac{2 \pi m}{p}|}.
\ea
\ee
Finally, the \oupb {\bf GenPyramid} is
\be
\vec{p}_i = \vec{v}_i \otimes  \vec{v}_{2i\, \bmod\, p}\otimes  \ldots
\otimes \vec{v}_{ni\, \bmod\, p},
\;\; i=0,\ldots,2n.
\label{upb3}
\ee
 
The primality of $p$ ensures that there are no states repeated on one
side: there is no $k$ in the range $1\leq k\leq 2n$ such that
$ki\,\bmod\,p=kj\,\bmod\,p$ for some integers $i \neq j$ if $p$ is
prime.  Orthogonality is also ensured by primality. As in
Fig. \ref{sept} there will be a party for whom next neighbor states
are orthogonal, there will be a party for whom all second neighbor
states are orthogonal, etc. up to the $n$th neighbor. This implies
that all vertices in the orthogonality graph are mutually connected
(orthogonal), so the orthogonality graph is complete. From basic
three dimensional geometry it follows that any set of three vectors
has full rank when $h_p \neq 0$ and thus these generalized sets form
\oupbs.
 
It was noted by A. Peres \cite{privperes} that this construction is
quite general: instead of the vectors of Eq. (\ref{defgensept}), we
take any set of vectors $\ket{r_i}$ such that $\langle r_i\, |\,
r_{i+1 \bmod p} \rangle=0$ and such that any triplet of vectors
$(\ket{r_i},\ket{r_k},\ket{r_l})$ with $i \neq j \neq l$ spans the full three
dimensional space. We construct the vectors $\vec{p}_i$ as in
Eq. (\ref{upb3}) with ${\vec v}_{mi\,\bmod\, p}={\vec r}_i$, with $m$
given in Eq. (\ref{boundm}).  This set can form a UPB.  But in this more
general construction a more complete check is required to make sure that any three
different vectors are linearly independent. If we restrict ourselves
to just requiring that adjacent vectors be orthogonal, we find that
there are sets with the same orthogonality graph as, for example,
{\bf Sept}, but which are not UPBs. An example of such a set is the
following:
\be
\ba{cccc}
\ket{r_1}=\left(\ba{r} 1 \\ 0 \\ 0 \ea \right), &
\ket{r_2  }=\left(\ba{r} 0 \\ 1 \\ 0 \ea \right), &
\ket{r_3}=\left(\ba{r} 1 \\ 0 \\ 1 \ea \right), &
\ket{r_4}=\left(\ba{r} 1 \\ 1 \\ -1 \ea \right), \\
& & & \\
\ket{r_5}=\left(\ba{r} 1 \\ -1 \\ 0 \ea \right), &
\ket{r_6}=\left(\ba{r} 1 \\ 1 \\ 1 \ea \right), &
\ket{r_7}=\left(\ba{r} 0 \\ 1 \\ -1 \ea \right). &
\ea
\ee
For these states we have $\langle r_i\, |\, r_{i+1 \bmod 7} \rangle=0$
and no other states are orthogonal. We can construct a PB by replacing
the states in Eq. (\ref{upb3}) by these vectors
$\vec{v}_{2i\,\bmod\, 7}=\vec{r}_i$.  This PB
has the same orthogonality graph as {\bf Sept}. However, the vectors
$\ket{r_1},\ket{r_2}$ and $\ket{r_5}$ lie in a two dimensional
plane. This implies that we can add a new product vector to the PB
thus constructed. This provides the counterexample to the idea that there is
a straightforward correspondence between orthogonality graphs and UPBs.
 
\subsection{UPBs based on quadratic residues}
\label{quadresi}

{\bf QuadRes} is a family of UPBs, which are based on
quadratic residues \cite{hardy&wright}.  The UPB is a set of orthogonal product
states on $n \otimes n$ where $n$ is such that $2n-1$ is a prime $p$ of the form
$4m+1$. The set contains $p=2n-1$ members, which is the minimal
number for a \oupb. Thus we can have
$(m,p,n)=(1,5,3),(3,13,7),(4,17,9)$, etc. The first triple $(1,5,3)$
is the {\bf Pyramid} UPB.
 
Let ${\bf Z}_p^*$ be ${\bf Z}_p \setminus \{0\}$.
Let $Q_p$ be a group of quadratic residues, that is, elements $q \in {\bf Z}_p^*$ such
that
\be
q=x^2 \bmod p,
\ee
for some integer $x$. $Q_p$ is a group under multiplication. The order
of the group is $\frac{p-1}{2}$. The following properties hold: when
$q_1 \in Q_p$ and $q_2 \notin Q_p$, a quadratic nonresidue, then $q_1
q_2 \notin Q_p$. Also, if $q_1 \notin Q_p$ and $q_2 \notin Q_p$, then
$q_1 q_2 \in Q_p$ \cite{hardy&wright}. The states of the UPB are
\be
\ba{lcr}
\ket{Q(a)}\otimes \ket{Q(xa)} & \mbox{ for } a \in {\bf Z}_p,  & x \in
{\bf Z}_p^*,\; x \not\in Q_p,
\ea
\label{resstates1}
\ee
where
\be
\ket{Q(a)}=(N,0,\ldots,0)+\sum_{q \in Q_p}e^{2 \pi i q a/p} \hat{e}_q,
\label{resstates2}
\ee
where $N$ is a real normalization constant to be fixed for
orthogonality and $\hat{e}_q$ are unit vectors of the form $(0,1,0,
\ldots, 0)$, $(0,0,1,\ldots,0)$ etc. The dimension $n$ of the Hilbert
space is $\frac{p+1}{2}$, one more than the order of $Q_p$.  One can
prove that these vectors can be made orthogonal by an appropriate
choice of $N$, for $a\neq b$:
\bea
\bra{Q(a)} Q(b)\rangle \bra{Q(xa)}Q(xb) \rangle= \nonumber \\
(|N|^2+\sum_{q \in Q_p}e^{2 \pi i q (b-a)/p})\,
(|N|^2+\sum_{q \in Q_p}e^{2 \pi i q x(b-a)/p})=0.
\label{ortho_res}
\eea
One uses the properties of $Q_p$ to find that for $b-a \neq 0$:
\be
\sum_{q \in Q_p}e^{2 \pi i q (b-a)/p}+\sum_{q \in Q_p} e^{2 \pi i q x(b-a)/p}=
\sum_{z \in {\bf Z}_p^*}e^{2 \pi i z (b-a)/p}=-1.\label{qrprop}
\ee
Thus the orthogonality relation of Eq. (\ref{ortho_res}) for $b \neq a$ is of the form
\be
(|N|^2+s)\,(|N|^2-1-s)=0,
\label{orthox}
\ee
where
\be
s=\sum_{q \in Q_p}e^{2 \pi i q (b-a)/p}.
\ee
 
The value of $s$ as a function of $b-a$ ($b \neq a$) only depends on whether $b-a \in Q_p$, because of the group property of $Q_p$. Call this value $\bar s$ when $b-a \in Q_p$. Then when
$b-a \not\in Q_p$, $s=-1-\bar{s}$ because of Eq. (\ref{qrprop}).
Finally we also need to establish that $s$ is real. One considers $s^*$ in which one sums over $-q$. As $q \in Q_p$ and
$-1 \in Q_p$ when $p$ is of the form $4m+1$ (see Theorem 82,
\cite{hardy&wright}), $-q \in Q_p$. Thus $s=s^*$. Both for negative as
well as positive $s$, Eq. (\ref{orthox}) has a solution for $N$,
and thus Eq. (\ref{ortho_res}) is satisfied for all $a\neq b$.
 
\begin{theo}
The states given in Eq. (\ref{resstates1}) and Eq. (\ref{resstates2}) on $n \otimes n$ with $2n-1$ a prime of
the form $4m+1$ with the appropriate value of $N$ determined by the
solution of Eq. (\ref{orthox}) form a UPB.
\label{conjpeter}
\end{theo}
 
\noindent {\it Proof}. The proof requires the application of Lemma \ref{rule1}, that is,
one must show that any set of $n$ states on either side spans the full
$n$-dimensional Hilbert space.  To do this, we need to show that any
subset of $n = (p+1)/2$ of the $p$ vectors defined in Eq.~(\ref{resstates2}) 
has full rank.  Checking whether a subset $T$ of these has full rank is
easily seen to be equivalent to checking whether the determinant of a
matrix $M_{Q_pT}$ does not vanish, where
$M_{Q_pT}$ is the $(p+1)/2 \times (p+1)/2$ matrix whose $j,k$ entry is 
$e^{2\pi i q_j t_k /p}$, $q_j$ being the $j$'th element of the set
$Q_p$ and $t_k$ the $k$'th element of the set $T$.  However, 
a theorem of \v Cebotarev \cite{cebotarev} shows that the matrix $M_{ST}$
is of full rank for any two arbitrary sets $S, T$, subsets of 
$\{0,1, \ldots, p-1\}$, proving Theorem \ref{conjpeter}. $\Box$

Drawn as orthogonality graphs as in
Fig. \ref{pentfig}, these UPBs are regular polygons, with a prime number 
$p$ (of the form
$4m+1$) of vertices. The elements of the quadratic residue group $Q_p$
correspond to the periodicity of the vectors that are orthogonal on one
side. For example, when $p=13$, one has quadratic residues $1,3,4,9,10$ and
$12$. Thus on, say, Alice's side, every vertex is connected to its first
neighbor (1), every vertex is connected with the $3$rd neighbor (3) etc.
On Bob's side the orthogonality pattern follows from the quadratic
nonresidues.

\subsection{Tensor powers of UPBs}
\label{tpupbsec}
 
When we have found two UPBs, we may ask whether the tensor product of them is
again a UPB. The answer is yes, as indicated by the following theorem:
 
\begin{theo} Given two bipartite UPBs ${\rm S}_1$ with members $\ket{\psi_i^1}$, $i=1,\ldots,l_1$ on
$n_1 \otimes m_1$ and ${\rm S}_2$ with members $\ket{\psi_i^2}$, $i=1,\ldots, l_2$ on $n_2 \otimes m_2$. The PB $\{\ket{\psi_i^1} \otimes \ket{\psi_j^2}\}_{i,j=1}^{l_1,l_2}$ is a bipartite UPB on $n_1 n_2 \otimes  m_1 m_2$.
\label{tpupb}
\end{theo}
 
\noindent{\it Proof}. Assume the contrary, i.e. there is a product
state that is orthogonal to this new ensemble which we call ${\rm
PB}^2$. The idea is to show that this leads to a contradiction and
thus PB$^2$ is a UPB.
 
Note first that for any \oupb a partition $P$ into a set with 0 states
for Bob and all states for Alice gives rise to a $r^P_A=\dim {\cal H}_A$ (see Lemma \ref{rule1});
the states on Alice's side together must span the entire Hilbert space
of Alice. Also note that if one takes a tensor product of two \oupbs
(defined on ${\cal H}_{A_1} \otimes \ldots$ and ${\cal H}_{A_2} \otimes \ldots$) this partition in which all states are assigned to Alice still leads to
$r^P_A=\dim {\cal H}_{A_1} \dim {\cal H}_{A_2}$.

\begin{figure}[htbf]
\centering
\epsfxsize=4.0in
\leavevmode\epsfbox{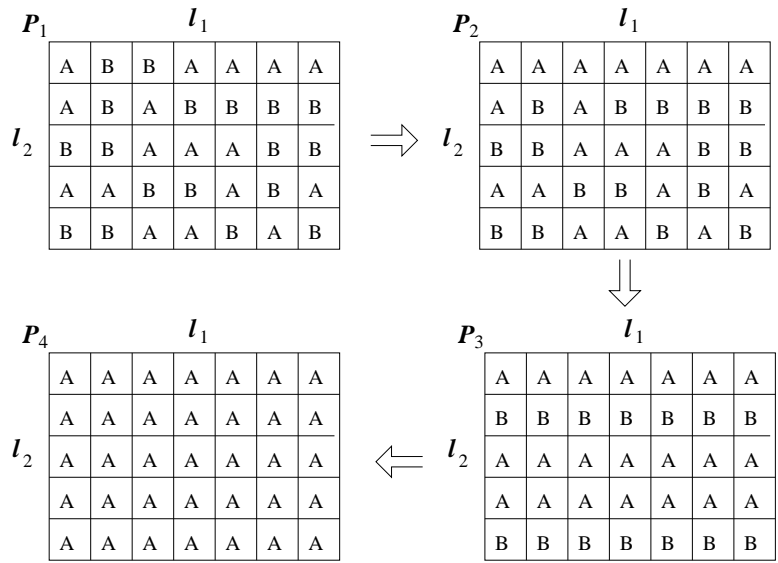}
\caption{The succession of partitions of the tensor power of UPBs used
to prove Theorem \protect\ref{tpupb}.  The As and Bs denote on what
side a hypothetical product state is orthogonal to the members of PB$^2$.}
\label{rect}
\end{figure}
 
The set PB$^2$ has $l_1 l_2$ members. The new hypothetical
product state $\Psi$ to be added to the set has to be orthogonal to
each member either on Bob's side or on Alice's side, or on both
sides. One can represent this hypothetical orthogonality pattern as a
rectangle of size $l_1$ (number of columns) by $l_2$ (number of rows)
filled with the letters A and B, depending on how the new state is
orthogonal to a member of the PB$^2$.  This is illustrated as
partition $P_1$ in Fig. \ref{rect}.  When the hypothetical state
$\Psi$ is orthogonal on both sides, we are free to choose an A or B in
the corresponding square.
 
Consider a row of this rectangle, for example the first one.  The
pattern of As and the Bs can be viewed as a partition of the ${\rm
S}_1$ \oupb and therefore one knows that one of these sets, either the
A or the B set, must have full local rank. But if the A set is the one
with full local rank, then the state $\Psi$ is also orthogonal to
PB$^2$ with respect to partition $P_2$ (Fig. \ref{rect}) in
which the whole row is filled with As. This is so since the local rank $r_A$
 is not changed (because the A rank was already full for the
row) and the rank $r_B$ cannot increase (since states are
removed from the B set). This is true for every row in turn, which leads us to partition $P_3$.
Then doing the same to columns, since all columns are identical they
will all be full rank for either A or B, so we will obtain the
unanimous partition $P_4$. This partition $P_4$ however contradicts the
fact that both sets were UPBs. $\Box$
 
The theorem has the consequence that arbitrary tensor powers of
bipartite UPBs are again UPBs. A generalization of the theorem for
multipartite states is true as well; in the proof we replace A/B partitions
with partitions of As, Bs, Cs etc.
 
\section{The Use of Bound Entanglement}
\label{usebe}
 
It has been shown \cite{teleBE} that bound entangled PPT states are
not a usefull resource in the teleportation of quantum states. On
the other hand it has also been shown \cite{BEactivate} that bound
entangled PPT states can have a catalytic effect in the (quasi) distillation of a
single entangled state. In the next two sections we discuss the use
of bound entangled states: in a protocol of distillation of mixed entangled
states and in defining a binding entanglement channel.

\subsection{Distillation of Mixed Entangled States}
 
We will prove that bound entangled states cannot be used to increase the
distillable entanglement of a state $\rho$ beyond its regularized entanglement of formation assisted by bound entanglement $E_b(\rho)$. By a bound
entangled state $\rho_b$ we mean a state that cannot be distilled,
i.e. if we are given many copies of this density matrix we cannot
distill {\em any} pure entanglement out of this set. These set of states
includes the bound entangled PPT states and also possibly some NPT
entangled states \cite{nptnond1,nptnond2}.

We denote the density matrix of a set of $n$ pure EPR pairs or other
maximally entangled states as $\Pi_{\rm EPR}^{\otimes n}$.  The
precise definition of distillable entanglement uses a limit in which
the number of copies $n$ of the state to be distilled goes to infinity
while at the same time the fidelity of the distilled states with
respect to a maximally entangled state goes to 1. In the notation we
use here we omit these limits for the sake of clarity. We refer the
reader to Ref. \cite{rainsdist} for a treatment and discussion of various
equivalent definitions of distillable entanglement.
 
We start with the following lemma:
 
\begin{lem}
For no integer $k$ and bound entangled state $\rho_b$ does there exists a 
{\rm LQ+CC TCP} map ${\cal S}_1$ such that
\be
{\cal S}_1(\Pi_{\rm EPR}^{\otimes n}\otimes\rho_b^{\otimes k})=
{\Pi^{\otimes 3n}_{\rm EPR}}.
\label{activ1}
\ee
\label{no3fac}
\end{lem}
 
\noindent{\it Proof}. Suppose Eq. (\ref{activ1}) were true.  Expanding
the density matrix proportional to the identity on the $2^n\otimes
2^n$ dimensional Hilbert space, we obtain \be {1\over 4^n}I={1\over
4^n}\Pi_{\rm EPR}^{\otimes n}+{4^n-1\over 4^n}\delta\rho.
\label{id}
\ee
By the linearity of ${\cal S}_1$ it follows that
\be
{\cal S}_1\left({1\over 4^n}I\otimes\rho_b^{\otimes k}\right)=
{1\over 4^n}{\Pi^{\otimes 3n}_{\rm EPR}}+
{4^n-1\over 4^n}{\cal S}_1(\delta\rho\otimes\rho_b^{\otimes k}).
\label{noe}
\ee
The fidelity of the output state in Eq. (\ref{noe})
with respect to $\Pi_{\rm EPR}^{\otimes 3n}$ is $F\geq 1/4^n$.  If the
output is projected into the Hilbert space of
dimension is $d\otimes d=2^{3n}\otimes 2^{3n}$ inhabited by
the $\Pi_{\rm EPR}^{\otimes 3n}$ term of Eq. (\ref{noe}) this fidelity
can only increase or remain the same. It has been shown by Horodecki {\em et al.} \cite{filterhor} that a state
for which $F>1/d$ is distillable, so the output state is distillable
(as $1/4^n > 1/2^{3n}$). But this is a contradiction, since the input state of Eq. (\ref{noe}) has only bound
entanglement, and the TCP map is LQ+CC and therefore it cannot create any free entanglement.  This proves that such an LQ+CC ${\cal S}_1$ does not
exist.$\Box$
 
\begin{lem}
For no integer $k$ and bound entangled state $\rho_b$ and $\alpha>1$ does 
there exists a {\rm LQ+CC TCP} map ${\cal S}_2$ such that
\be
{\cal S}_2(\Pi_{\rm EPR}^{\otimes n}\otimes\rho_b^{\otimes k})=
\Pi_{\rm EPR}^{\otimes \alpha n}.
\label{activ2}
\ee
\label{noafac}
\end{lem}
 
\noindent{\it Proof}. If ${\cal S}_2$ existed, iterated application of it
${\cal S}_2({\cal S}_2({\cal S}_2(...(\Pi_{\rm EPR}^
{\otimes n}\otimes\rho_b^{\otimes k})...)))$ $\log 3/\log\alpha$ times would
produce the map ${\cal S}_1$ of Lemma \ref{no3fac}. However this ${\cal S}_1$ cannot
exist and therefore ${\cal S}_2$ does not exist. $\Box$
 
The distillable entanglement of a state $\rho$ assisted
by bound entanglement, $D_b(\rho)$, is defined by optimizing over
all LQ+CC TCP maps and bound-entangled states $\rho_b$ and values for $k$
such that
\be
{\cal S}_3(\rho^{\otimes n}\otimes\rho_b^{\otimes k})=
\Pi_{\rm EPR}^{\otimes D_bn},
\label{activ3}
\ee
 
\begin{propo} $D(\rho) \leq  D_b(\rho)\leq E_b(\rho) \leq E(\rho)$, where ($E_b(\rho)$) $E(\rho)$ is (the $BE$-assisted) regularized entanglement of formation of $\rho$.
\end{propo}
 
\noindent {\it Proof}. The BE-assisted regularized entanglement of formation $E_b(\rho)$ of a density matrix $\rho$ is defined by the optimal LQ+CC TCP map ${\cal S}_{E_b}$ and optimal choice for $k$ and
$\rho_b$ such that
\be
{\cal S}_{E_b}(\Pi_{\rm EPR}^{\otimes E_b n} \otimes \rho_b^{\otimes k})=\rho^{\otimes n}.
\label{activE}
\ee
Suppose $D_b(\rho)>E_b(\rho)$.  This leads to a contradiction, because
the composed map
\be
{\cal S}_3({\cal S}_{E_b}(\Pi_{\rm EPR}^{\otimes E_b n} \otimes \rho_b^{\otimes k})\otimes\rho_b^{\otimes l})=\Pi_{\rm EPR}^{\otimes D_bn},
\ee
cannot exist by Lemma \ref{noafac}. $\Box$
 
These results also provide some partial
answers to the questions raised in the discussion of Ref. \cite{bind}; it bounds
the use that bound entanglement can have in the distillation of mixed states.
The result does leave room for nonadditivity though; for states which have
$D(\rho) < E(\rho)$ it could still be that $D(\rho) < D_b(\rho)$.

\subsection{Binding Entanglement Channels}
\label{binding}
 
As noted independently by Horodecki {\em et al.} \cite{bind}, there
exist quantum channels through which entanglement can be shared, but
only entanglement of the bound variety.  These {\em binding entanglement
channels} are discussed in Ref. \cite{bind}.  Here
we present a simple physical argument for their existence
based on bound entangled states, both of the PPT kind as well as the
NPT kind (if these exist, see Ref. \cite{nptnond1}).
 
\begin{figure}
\centering
\epsfxsize=3.5in
\leavevmode\epsfbox{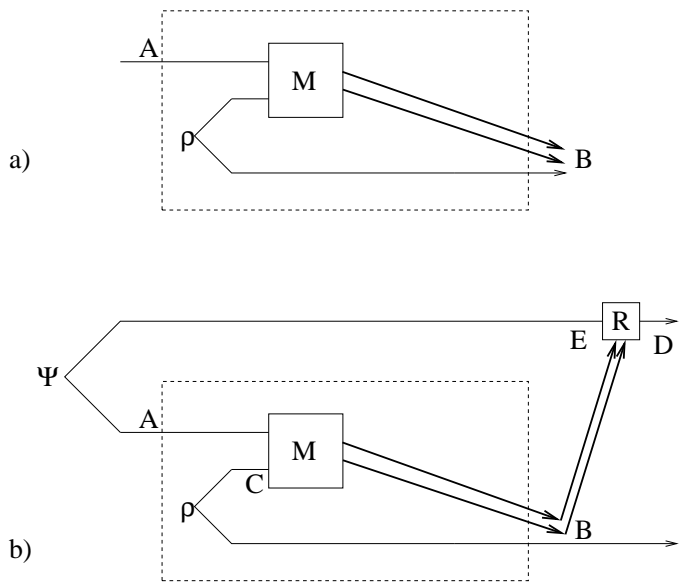}
\caption{Binding Entanglement Channels: a) The input (from point A) is
measured along with half of the BE state $\rho$ in a maximally entangled
basis. The measurement M produces classical information represented by
the heavy lines.  The classical results along with the other half of
$\rho$ are returned at the output B.  b) Alice sends half of a
maximally entangled pair $\Psi$ through the channel.  Bob sends the
classical data back to Alice, who then performs rotation R on the
remaing half of $\Psi$ as determined by the data.  The result is
teleportation from C to D.}
\label{bindingfig}
\end{figure}

Consider any bound entangled (BE) state $\rho$ on $m\otimes m$. With this
state we define a channel which
takes an $m$-dimensional input and measures it, along with
one half of $\rho$, in a basis of maximally entangled states.  The
output of the channel is the other half of $\rho$ and the classical
result of the measurement, see Fig. \ref{bindingfig}(a).
It is easy to see that no pure entanglement can ever be shared
through such a channel, as any procedure which could would
also be able to distill entanglement from the BE state $\rho$ itself.
No pure entanglement can ever be shared through such a channel, since 
any procedure which could would also be able to distill entanglement 
from the BE state $\rho$ itself: If Alice and Bob share many copies 
of $\rho$, they can simulate actually having the channel by having Alice 
measure each of her inputs to the channel along with her half of a copy of $\rho$ in the basis of maximally entangled states, and telling Bob the classical result, just as the channel itself would have done.  By plugging their simulated
channel into a procedure that could share pure entanglement through
the channel, Alice and Bob would have distilled entanglement from
the bound entangled state $\rho$.

This does not yet establish the existence of a BE channel as
our channel might only be able to share separable states.
Now suppose Alice, whose lab is at the top of Fig. \ref{bindingfig}(b),
creates a maximally entangled state $\Psi$ in $m\otimes m$ and
sends half of it into the channel.  Bob, whose lab is at the bottom
right of the figure, sends the classical output data to Alice,
who does some unitary operation R depending on those data.  If the
set of possible R's is chosen correctly, the result is precisely
quantum teleportation \cite{tele} of the half of $\rho$ at
point C to point D.  So a bound entangled state has been shared
through the channel.  Finally, we note that the actual transmission
of the classical data from Bob to Alice, while a simplifying idea,
is not strictly needed.  That communication along with rotation R
is a LQ+CC operation and therefore cannot create entanglement
where there was none.  So even before the classical communication
the state shared between points B and E must have been bound
entangled.

\section{Conclusion}
We have shown some of the mathematical richness of the concept of
unextendible and uncompletable product bases, their relation to graph theory
and number theory. By exhibiting some of this structure we have uncovered
a large family of bound entangled states. We have presented the first example
of a new construction for bound entangled states. It would be interesting
to try to understand the geometry of uncompletable product basis in a
more general way; some of the interesting open questions in this respect have
been mentioned in the paper. For example, from every multipartite UPB we
can derive bipartite PBs by considering the UPB over bipartite cuts. Do
these PBs have any special properties; can they correspond UCPBs when
the local Hilbert spaces have dimension more than 2?
 
The question that this work
only partially addresses is one concerning the fruitful use of bound entangled
states and the resources needed to implement separable superoperators.
Further investigations into this matter will be worthwhile.

{\bf Acknowledgements}. Part of this work was completed during the 1998 Elsag-Bailey-I.S.I. Foundation
research meeting on quantum computation. We would like to thank
Charles Bennett, Asher Peres, Danny Terno, Ashish Thapliyal and John Tromp 
for discussion. We would like to thank Noga Alon for bringing \v Cebotarev's
theorem to our attention, thus proving Theorem \ref{conjpeter}. 
JAS and DPD acknowledge support from the Army Research Office under contract 
number DAAG55-98-C-0041. The work of TM was supported in part
by grant \#961360 from the Jet Propulsion Lab, and grant \#530-1415-01 
from the DARPA Ultra program.

\appendix
 
\section{No six member UPB in $3 \otimes 3$}
\label{nosixmem}
 
\begin{figure}[htbf]
\centering
\epsfxsize=4.0in
\leavevmode\epsfbox{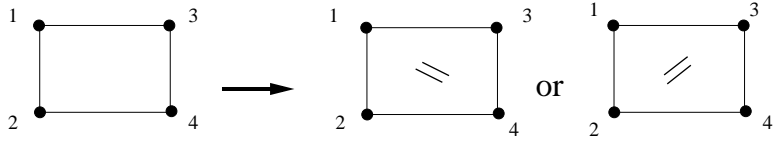}
\caption{The square rule for an orthogonality graph in $3 \otimes 3$.}
\label{rules}
\end{figure}

\begin{figure}[htbf]
\centering
\epsfxsize=4.0in
\leavevmode\epsfbox{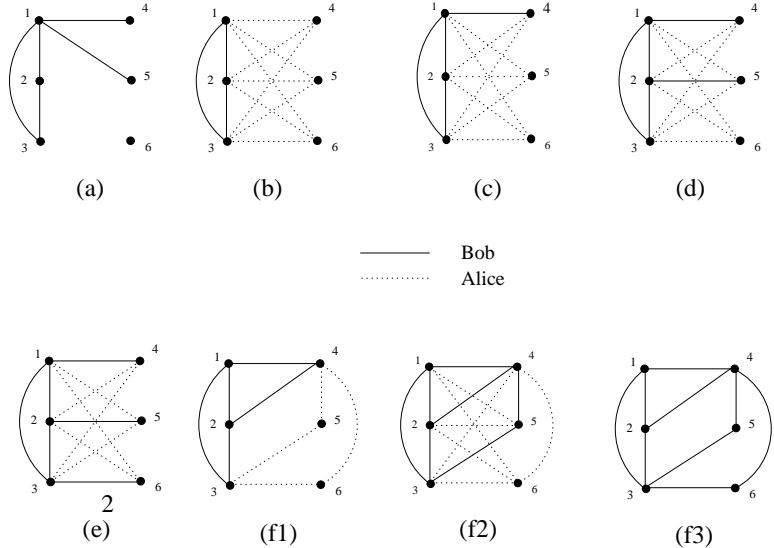}
\caption{The orthogonality graphs of PBs with six
members on $3 \otimes 3$.}
\label{graphs6}
\end{figure}

In this appendix we prove that there cannot exist a UPB with six members in
$3 \otimes 3$. We use some elementary graph theory to simplify the argument. We denote the
complete graph on $n$ vertices as $K_n$, i.e. in this graph all pairs of vertices
are connected by an edge. The Ramsey number $R(s,t)$ (cf. Ref. \cite{bollobas})
is defined as the smallest number $n$ such that every coloring of the edges
of $K_n$ with 2 colors, say red and blue, contains either a red $K_s$ or a blue $K_t$. The Ramsey number $R(3,3)=6$. This implies that the graph of any
product basis with six members contains at least three states which are mutually orthogonal either on Alice's or Bob's side ; they form an orthogonal
triad. Let us assume that this occurs on Bob's side. We label these
states as $\ket{\beta_1},\ket{\beta_2},\ket{\beta_3}$. Before considering
some special cases we establish a simple rule which follows from the fact that the
states are defined on $3 \otimes 3$; it is depicted in graph language
in Fig. \ref{rules}. Fig. \ref{rules} says then when we have a connected
square of one color, there will be a repeated state, denoted by the equality
``$=$'' sign. Let $\ket{\beta_1}=\ket{1}$ and $\ket{\beta_2}=\ket{0}$, then
$\ket{\beta_3}=\ket{1^{\perp}} \in {\rm span}(\ket{0},\ket{2})$ and $\ket{\beta_4}=\ket{0^{\perp}} \in {\rm span}(\ket{1},\ket{2})$. Orthogonality of $\ket{\beta_3}$ and
$\ket{\beta_4}$ implies that either $\ket{\beta_3}=\ket{0}$ or $\ket{\beta_4}=\ket{1}$.
Now we consider some subcases. In these cases the non UPB character of the
set is derived, either by directly showing how to extend the set or by
showing that the states can be distinguished by LQ+CC (see Lemma
\ref{immeasurability}).
We have depicted the cases in Fig. \ref{graphs6}:\\
(a) There exists an $\ket{\beta_i} \in \{\ket{\beta_1},\ket{\beta_2},\ket{\beta_3}\}$ such that this vertex $i$ is connected to two out of $\ket{\beta_{4,5,6}}$, say $\ket{\beta_4},\ket{\beta_5}$ on Bob's side. Then state
$\ket{(\alpha_i,\alpha_6)^{\perp}} \otimes \ket{\beta_i}$ is orthogonal to
all the members of the PB and thus the PB is extendible. \\
(b) None of the states $\ket{\beta_i}$ is orthogonal to any of
$\ket{\beta_{4,5,6}}$; then Alice can perform a dissection of the set
into $(1,2,3)$ and $(4,5,6)$. Proposition \ref{oneprime} then applies. \\
(c) There exists one state $\ket{\beta_i} \in \{\ket{\beta_1},\ket{\beta_2},\ket{\beta_3}\}$ such that this vertex $i$ is connected to exactly one out of $\ket{\beta_{4,5,6}}$ on Bob's side. For example, $i=1$. This means that Alice can do a von
Neumann measurement with $\Pi_{{\rm span}(\alpha_2,\alpha_3)}$ and $\Pi_{{\rm span}(\alpha_4,\alpha_5,\alpha_6)}$.
This will split the state $\ket{\alpha_1}$, but as we have seen before a von Neumann measurement that cuts a single state is orthogonality preserving. After the measurement
three or four orthogonal states are left to be distinguished. They can be
distinguished (Proposition \ref{oneprime} and Theorem \ref{no34}) and thus all six states can be distinguished. \\
(d) Here we consider the case in which two vertices, say, $\ket{\beta_1}$ and
  $\ket{\beta_2}$ are connected to two different vertices out of
$\ket{\beta_{4,5,6}}$. Notice that there is a square on the vertices
2,4,3,6 on Alice's side. This implies (see Fig. \ref{rules}) that either
2 is equal to 3 on Alice's side, which implies that 2 is also orthogonal to 5
on Alice's side, which results in case (c), or 4 is equal to 6 on Alice's
side which implies that 4 is orthogonal to 1 on Alice' side which also
results in a variant of case (c). \\
(e) Here we consider the case in which all three vertices $\ket{\beta_{1,2,3}}$ are connected to the three different vertices $\ket{\beta_{4,5,6}}$. When we try to connect, say, vertices 4 and 5 on Bob's side, we create a square and
extra orthogonalities, such that we find examples of case (a) on Bob's side.
If we connect all three vertices 4,5,6 on Alice's side, we get examples of
case (a) on Alice's side.\\
(f) When two vertices, say 1 and 2, are connected to the same vertex, say 4, on
Bob's side, it must be that state 3 is equal to state 4 on Bob's side.
Then there are three subcases. In case (f1) state 3 and therefore 4 is
not connected to 5 or 6 on Bob's side. Let us consider
how we can connect 1 to 5 and 2 to 5. With any choices of coloring of these
edges we create examples of case (a) on either Alice's or Bob's side.
In case (f2) 3 and therefore 4 is only connected to, say, state 5 on
Bob's side. Then to avoid case (a) on Bob's side we put Alice's
edges between 1 and $(5,6)$ and 2 and $(5,6)$ and 3 and 4. But then a case (a)
occurs on Alice's side. In case (f3) both 3 or 4 are connected to 5 and 6 on Bob's side; this creates a case (a) again on Bob's side.
 
This establishes the no-go result for a $6$-member UPB in $3 \otimes 3$.
 
\bibliographystyle{hunsrt}
\bibliography{refs}

\begin{thebibliography}{10}
\bibitem{fnt1}
This matrix transposition can be carried out in any basis. The resulting
  matrices that one obtains by transposing in different bases are identical up
  to a local unitary transformation and therefore they have identical
  eigenvalues which determine whether they are positive semidefinite or not.
\end{thebibliography}

\end{document}